\documentclass[traditabstract]{aa}

\usepackage{color}
\usepackage{graphicx}
\usepackage{txfonts}
\include{epsf}
\usepackage{natbib}
\usepackage{url}
\usepackage{fixltx2e}
\bibpunct{(}{)}{;}{a}{}{,}

\def\0{\phantom0}

\def\kms{km s$^{-1}$}

\def\ediscs{EDisCS}

\begin{document}

\title{Disc colours in field and cluster spiral galaxies at $0.5 \lesssim z \lesssim 0.8$}
\titlerunning{EDisCS: Discs colours in field and cluster spiral galaxies}

\author{Nicolas Cantale\inst{1} 
\and  Pascale Jablonka\inst{1,2} 
\and  Fr\'ed\'eric Courbin\inst{1}
\and  Gregory Rudnick\inst{3}
\and  Dennis Zaritsky\inst{4}
\and  Georges Meylan\inst{1}
\and  Vandana Desai\inst{5}
\and  Gabriella De Lucia\inst{6}
\and  Alfonson Arag\'on-Salamanca\inst{7}
\and  Bianca M. Poggianti\inst{8}
\and  Rose Finn\inst{9}
\and  Luc Simard\inst{10}
}
\institute{Laboratoire d'astrophysique, Ecole Polytechnique F\'ed\'erale de Lausanne (EPFL), Observatoire de Sauverny, CH-1290 Versoix, Switzerland 
\and  GEPI, Observatoire de Paris, CNRS UMR 8111, Universit\'e Paris Diderot, 92125, Meudon Cedex, France 
\and University of Kansas, Dept. of Physics and Astronomy, Lawrence, Kensas, USA 
\and Steward Observatory, University of Arizona, 933 North Cherry Avenue, Tucson, AZ 85721, USA 
\and {\it Spitzer} Science Center, Caltech, Pasadena CA 91125, USA 
\and INAF-Astronomical Observatory of Trieste, via G.B. Tiepolo 11, I-34143 Trieste, Italy 
\and School of Physics and Astronomy, University of Nottingham, University Park, Nottingham NG7 2RD
\and Osservatorio Astronomico di Padova, Vicolo dell'Osservatorio 5, 35122 Padova, Italy  (Bianca)
\and Department of Physics and Astronomy, Siena College, 515 Loudon Road, Loudonville, NY 12211, USA  
\and Herzberg Institute of Astrophysics, National Research Council of Canada, 5071 West Saanich Road, Victoria, Canada BC V9E 2E7
}

\date{Received; accepted }
 
\abstract{We present a detailed study of the colours in late-type
  galaxy discs for ten of the \ediscs\ galaxy clusters with $0.5 \lesssim z
  \lesssim 0.8$. Our cluster sample contains 172 spiral galaxies, and our
  control sample is composed of 96 field disc galaxies. We deconvolved their
  ground-based $V$ and $I$ images obtained with FORS2 at the
VLT with initial spatial resolutions
  between 0.4 and 0.8 arcsec to achieve a final resolution of 0.1 arcsec with 0.05
  arcsec pixels, which is close to the resolution of the ACS
at the HST.  After removing the
  central region of each galaxy to avoid pollution by the bulges, we
  measured the $V-I$ colours of the discs.  We find that 50\% of cluster spiral
  galaxies have disc $V-I$ colours redder by more than 1$\sigma$ of the mean
  colours of their field counterparts. This is well above the 16\% expected for a
  normal distribution centred on the field disc properties. The prominence of
  galaxies with red discs depends neither on the mass of their parent cluster nor
  on the distance of the galaxies to the cluster cores. Passive spiral galaxies
  constitute 20\% of our sample. These systems are not abnormally dusty. They are
  are made of old stars and are located on the cluster red sequences. Another 24\%
  of our sample is composed of galaxies that are still active and star forming,
  but less so than galaxies with similar morphologies in the field. These galaxies
  are naturally located in the blue sequence of their parent cluster
  colour-magnitude diagrams. The reddest of the discs in clusters must have stopped
  forming stars more than $\sim$ 5 Gyr ago. Some of them are found among
  infalling galaxies, suggesting preprocessing. Our results confirm that galaxies
  are able to continue forming stars for some significant period of time after
  being accreted into clusters, and suggest that star formation can decline on
  seemingly long (1 to 5 Gyr) timescales.}

\maketitle

\section{Introduction}
\label{intro}

Galaxy evolution is influenced by their
  environment \citep[for a review of the seminal works see][]{Blanton2009}, although
  the exact mechanisms at play and their sphere of influence still need to be
  unveiled \citep{Villalobos2012, Haines2015, Woo2015, Bahe2015}.  There are indeed
  spectacular signs of perturbations induced by the infall of galaxies in
  clusters, such as the stripping of the neutral gas \citep[e.g.][]{Chung2009}
  and complex perturbed morphologies \citep[e.g.][]{Ebeling2014}. Some
  signs are more subtle but nevertheless involve a large portion of
  a galaxy population, such as the decrease in star formation rate
  \citep[e.g.][]{Poggianti2008, Fumagalli2008}
or the emergence of lenticular galaxies (S0) and 
corresponding decline in spirals \citep[e.g.][]{Desai2007, Just2010}.
Although spiral disc fading is not a sufficient explanation for the rise of S0s
\citep{Christlein2004},  spirals are strongly
affected both in numbers and properties by dense environments. Clues to the intensity and timescale
of the processes at play are to be found in gas content and stellar population
properties \citep[e.g.][]{Delpino2014}.

The link between the hot gas, which is heated by young hot stars, and the galaxy stellar
component has essentially been investigated from the standpoint of their spatial
distribution. At intermediate redshift, \citet{Bamford2007} showed that while
the stellar disc scale lengths remain similar in field and cluster spirals, the
extent of the emission line regions is smaller in cluster systems. 
\citet{Jaffe2011} found that kinematically disturbed star-forming galaxies are
more frequent in clusters than in the field, but the presence of kinematically disturbed gas and morphological distortions are
almost not correlated at all.
They also showed evidence that the gas discs in cluster galaxies are truncated,
with star formation more concentrated than in a low-density environment. These
results are in line with those of \citet{Maltby2012}, who found with STAGES (the Space
Telescope A901/2 Galaxy Evolution Survey, \citep{Gray2009}) that the galaxy
environment does not affect the stellar distribution of the outer discs, but contrasts with studies in higher density environments such as the Coma and
Virgo clusters. In these, the stellar scale lengths of the disc galaxies were
measured to be 20-30\% smaller than those in the field
\citep{Aguerri2004,Gutierrez2004, Koopmann2006}.

The stellar population properties make a decomposition into bulge
and disc galaxies
necessary.  \citet{Hudson2010} analysed eight nearby clusters ($z \le0.06$, $700 <
\sigma < 1000$ \kms) and inferred that while the $B-R$ colours of the bulge
components do not significantly depend on the environment, the discs in the
cluster centre are redder than those at the virial radius and even more redder
than the discs of the field spirals. This was also confirmed by \citet{Head2014}
in the Coma cluster. Our work is part of a global observational effort to
  help identify how spirals are transformed in dense regions. Establishing where
  and when the stellar populations of cluster spiral discs differ from those in
  field analogues provides critical information on the gas-removal timescales
  for the bulk of the cluster galaxy population. To do this, we place our study at
  intermediate redshift, when both star formation and quenching are still in full swing \citep{Rudnick2009, Finn2010}

Distinguishing bulge and disc properties at intermediate and high redshifts
requires at least spatial resolutions and samplings comparable with those achieved
with the HST in at least two photometric bands. This type of dataset is not
always available, particularly in the case of large ground-based surveys. The
deconvolution technique offers a unique alternative. It is very efficient when
applied to objects with scales close to the point spread function (PSF) size and
presents a number of crucial advantages. First, the absence of priors on the
underlying galaxy profiles, classically imposed with analytical formulae, is expected to
provide more robust results. Second, because the PSF is essentially removed, the
resolution of the deconvolved images can be controlled and fixed for an entire
sample  despite very different original observing conditions. Third, the
improved sampling, combined with the higher resolution and the PSF removal, allows studying those parts of the galaxies that
were previously buried inside the
PSF.  The better the sampling of the initial PSF, the finer the results of the
deconvolution.

In the following, we present the first analysis of spiral disc
colours in
a sample of intermediate-redshift cluster galaxies by applying our deconvolution
technique to a set of images obtained with the FORS2 camera at
the VLT. Our sample is drawn from the ESO Distant
Cluster Survey (EDisCS, \cite{White2005}) and allows us to investigate trends with
cluster masses and lookback-time. Because the EDisCS clusters were only observed
with the ACS at the HST in the F814W filter, our deconvolution method provides a
unique opportunity of retrieving the galaxy properties at high spatial resolution
from the ground-based FORS2 multi-band observations.

We present our dataset in Sect. \ref{data}. An overview of the deconvolution
procedure is provided in Sect. \ref{deconv}. Section \ref{results} presents our
main results, which are discussed in Sect. \ref{discussion} and summarized in
Sect. \ref{conclusion}.

\section{EDisCS cluster sample}
\label{data}

\ediscs\ is an ESO Large Programme dedicated to the analysis of 20 galaxy clusters
in the redshift range $0.4\lesssim z \lesssim 1$, spanning a wide range of
masses, that is, of velocity dispersions, from $\sim$ 200 \kms\ to $\sim$ 1000
\kms\ \citep{Halliday2004, Milvang-JensenB.2008}. These clusters were initially
selected from the most significant brightness peaks in the Las Campanas Distant
Cluster Survey (LCDCS; \cite{Gonzalez2001}). They all benefit from deep 
$B$, $V$, $R$, and $I$ photometry obtained with FORS2 at the
VLT \citep{White2005}.  A subset of ten clusters was
also imaged with the ACS at the HST in the F814W filter \citep{White2005,Desai2007}; they
constitute the basis of the present study.

For the purpose of our analysis, we distinguish cluster and field populations
based on the cluster redshift, $z_{c\rm l}$, and cluster velocity dispersion, $\sigma_{\rm cl}$.
The cluster members are the galaxies with spectroscopic redshifts within $\pm 3
\sigma_{\rm cl}$ \citep{Halliday2004, Milvang-JensenB.2008}.  This spectroscopic
sample is magnitude limited at $I \sim23$ mag measured in 1 arcsec apertures, with the
exception of cl1232.5-1250, at lower redshift, for which the limit is  $I=22$ mag
\citep{Halliday2004}. As discussed in \cite{Poggianti2006}, we were able to build an
$I$-band selected unbiased sample of field galaxies with the same exact
observational conditions as the cluster members by selecting galaxies that are not cluster members in each observed
spectroscopic mask, but have redshifts within $\delta z \pm 0.1$ of the targeted cluster redshift.  

From the original sample of 350 spectroscopically confirmed cluster members, we
selected the spiral galaxies with T-type between 1 and 7, that is, with Hubble
types between Sa/SBa and Sd/SBd. The final set of galaxy members comprises 172
cluster members.  The full control sample encompasses 196 field galaxies, from
which 96 have Sa/SBa to Sd/SBd types and redshifts $0.45 \le z_{\rm spec} < 0.81$.
Table~\ref{tab:morph} summarises the number of cluster and field galaxies found in
each cluster field as explained above together with their morphologies. They constitute our
full sample, which is grouped into redshift
bins later in the analysis.

\begin{figure}[t!]
\centering
\includegraphics[width=9.0cm]{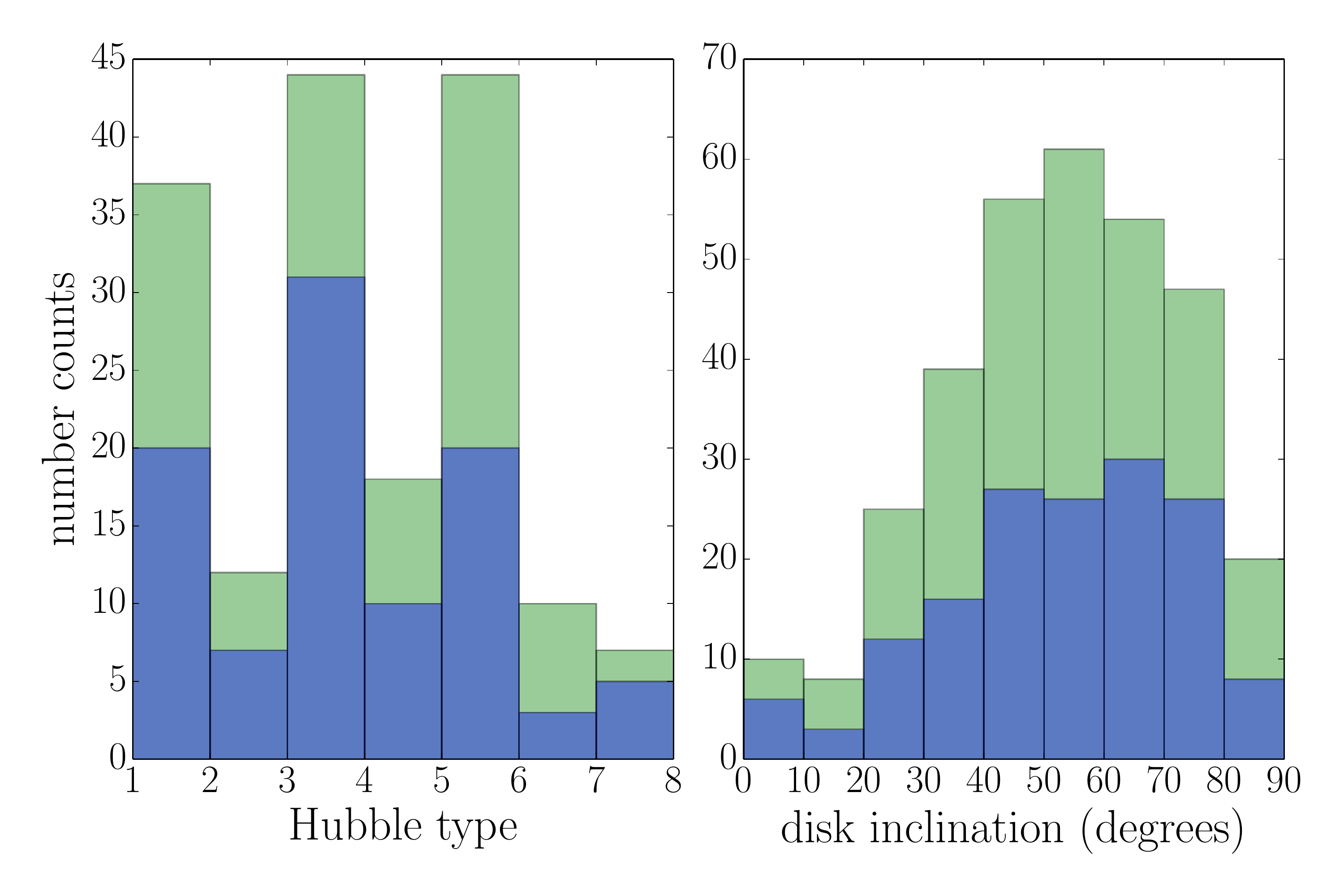}
\caption{Comparison of the distribution into Hubble types and inclinations between our
  cluster and field galaxy samples.  The cluster and field galaxies are plotted in
  green and blue, respectively.}
\label{fig:population}
\end{figure}

Figure~\ref{fig:population} displays the visually classified morphological types
\citep{Desai2007} of our sample galaxies and the distribution of the inclination
angles as measured from GIM2D \citep{ Simard2009}. The uncertainty on the
morphological classification is of the order of one Hubble type. The cluster
members and field galaxies display almost identical distributions and can thus
be safely compared free of biases that might have been generated by
differential dust obscuration or different morphologies.

\begin{table*}[!h]
\caption{Properties of the galaxy clusters: cluster name,
  redshift ($z_{\rm cl}$), velocity dispersion ($\sigma_{\rm cl}$), radius of the
  region covered by the observations in units of R$_{200}$ ($\mathrm{R_{\rm
      max}}$), number of field spiral galaxies defined as non-cluster member, but whose redshift falls within $\Delta z \leq 0.1$ of $z_{\rm cl}$
  ($N_{\rm field}$), number of cluster galaxies per Hubble type.}
\label{tab:morph}
\renewcommand{\arraystretch}{1.25} 
\setlength{\tabcolsep}{8pt} 
\begin{center}
\begin{tabular}{lccccccccc}
\hline
Cluster ID & $z_{cl}$ & \multicolumn{1}{c}{$\sigma_{\rm cl}$ [km$\,$s$^{-1}$]} & $\mathrm{R_{\rm max}/R_{200}}$ & $N_{\rm clus}$ & $N_{\rm field}$ & Sa-Sab & Sb-Sbc & Sc-Scd & Sd\\
\hline
cl1037.9-1243  & 0.5805 & $ 319^{~+53}_{~-52}$ & 1.99 & 11 & 14 & 3 & 6 & 1 & 1 \\
cl1040.7-1155 & 0.7043 & $ 418^{~+55}_{~-46}$ & 1.84 & 19 & 24 & 4 & 7 & 7 & 1 \\
cl1054.4-1146 & 0.6972 & $ 589^{~+78}_{~-70}$ & 1.85 & 20 & 12 & 5 & 9 & 5 & 1 \\
cl1054.7-1245 & 0.7502 & $ 504^{~+113}_{~-65}$ & 1.49 & 13 & 8 & 7 & 3 & 3 & 0 \\
cl1103.7-1245b & 0.7031 & $ 252^{~+65}_{~-85}$ & 2.71 & 6 & 11 & 0 & 3 & 3 & 0 \\
cl1138.2-1133  & 0.4788 & $ 732^{~+72}_{~-76}$ & 0.61 & 26 & 3 & 6 & 4 & 14 & 2 \\
cl1216.8-1201 & 0.7955 & $ 1018^{~+73}_{~-77}$ & 0.96 & 28 & 3 & 7 & 6 & 13 & 2\\
cl1227.9-1138  & 0.6357 & $ 574^{~+72}_{~-75}$ & 1.36 & 13 & 12 & 4 & 7 & 2 & 0 \\
cl1232.5-1250 & 0.5419 & $ 1080^{~+119}_{~-89}$ & 0.71 & 26 & 3 & 12 & 12 & 2 & 0 \\
cl1354.2-1230  & 0.7620 & $ 648^{+105}_{-110}$ & 1.59 & 10 & 6 & 1 & 5 & 4 & 0 \\
\hline
\end{tabular}
\end{center}
\end{table*}

\begin{figure*}[h]
\centering
\includegraphics[width=9cm]{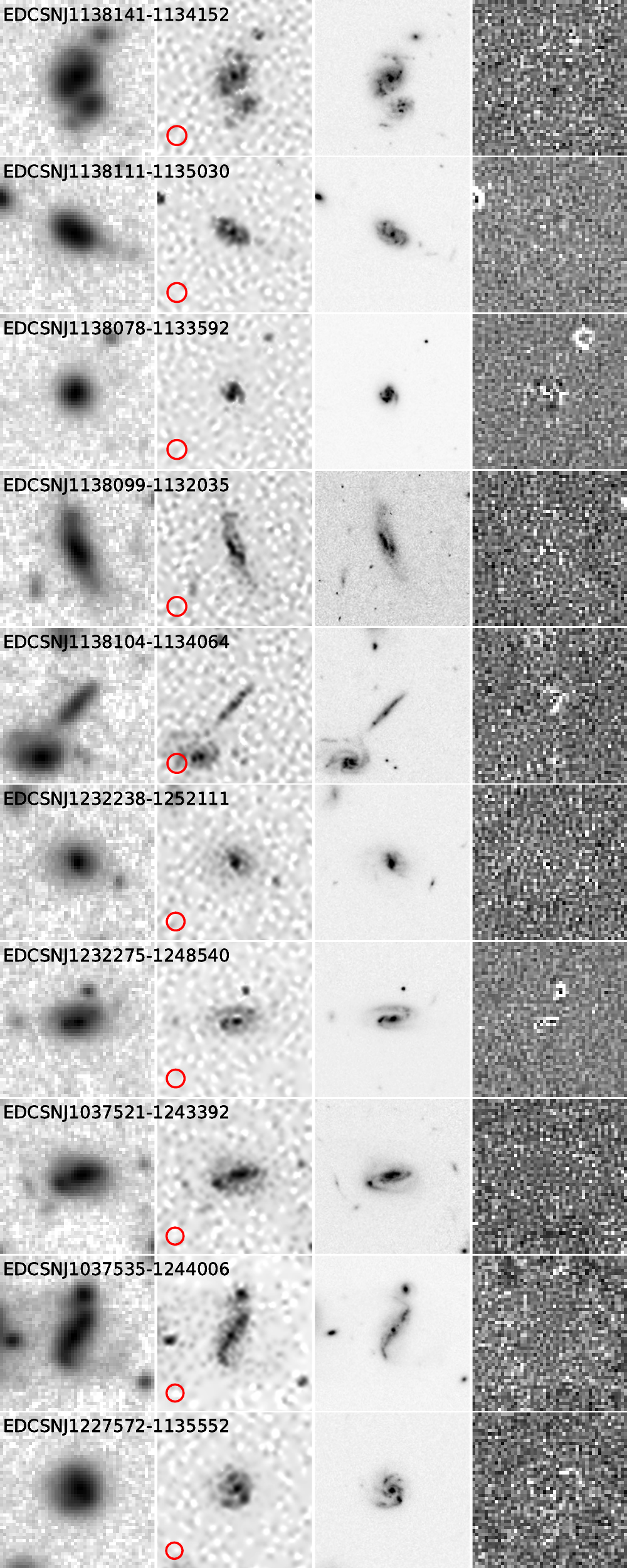}
\hskip 6pt
\includegraphics[width=9cm]{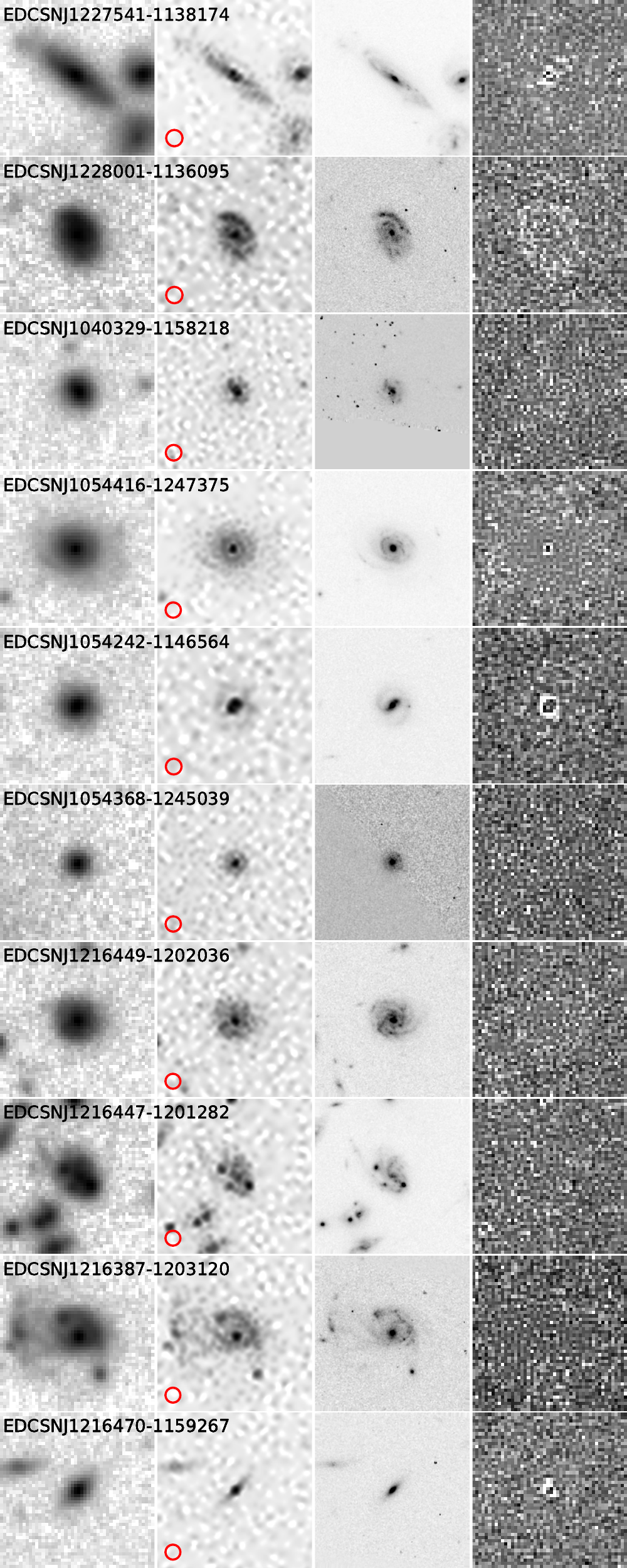}
\caption{Examples of deconvolving the \ediscs\ spiral galaxies over the whole
  redshift range. From left to right, the columns show (i) the original $I$-band images obtained with FORS2 at the VLT with
  seeing between 0.5\arcsec\ and 0.7\arcsec, (ii) their corresponding deconvolved VLT images with a
  resolution of 0.1\arcsec, (iii) the images obtained with the
ACS at the HST in the F814W filter for comparison, and (iv) the residual
  images in units of the photon noise between the initial FORS2 images and their deconvolution. The red
  circles represent the 3.5 kpc circular aperture used to remove the bulge
  contribution. The similarities between the VLT deconvolutions and the HST images
  are striking: details such as spiral arms or dust lanes previously smeared out by the
  PSF are made visible and are confirmed by the HST images.}
\label{Demo}
\end{figure*}

\section{Photometry}
\label{deconv}
\subsection{Image deconvolution}

Most cluster galaxies are small at $0.5 \lesssim z \lesssim 0.8$ . Spiral
bulges, whose effective radii are of the order of 2-3 ACS pixels, are therefore
barely resolved even with the HST.  The PSF full width at half
maximum (FWHM) of images obtained with FORS2 at the VLT lies
between 0.48\arcsec\ and 0.85\arcsec, depending on the photometric band and the
observing night.  The complete photometric description, with the seeings and
exposure times of each cluster, is accessible in \cite{White2005}. These resolutions
cause the bulge light to leak onto the disc component and adjust all
images to the same resolution, as is classically done in photometric
measurements. which further degrades the situation and makes studying the discs
impossible.

To distinguish the contribution of the different
galaxy components to the total luminosity, multi-component model fitting is often carried out. The impact
of the PSF is taken into account to deliver PSF-free galaxy
models as well as possible \citep[e.g.][]{Simard2002, Peng2011, Vika2013}. Nevertheless, the choice of
the analytical functions is somewhat arbitrary and does not necessarily apply to
all galaxies. The case of galaxies perturbed by gravitational interactions
constitute a clear problem for which a general analytical solution is not trivial, if it is possible at all. Therefore, we aim at avoiding any prior on the shape of the
galaxy components. The first step is to access the galaxy morphologies directly,
that is, visually. This is done by converting the original FORS2 images to a common and
much smaller resolution with a finer sampling similar to that of the HST. The
photometric measurements are then performed on these deconvolved images.  In the
following the HST images are only used as a posteriori checks of the reliability
of the deconvolution. They are not used as priors.

We used an improved version of the MCS image deconvolution algorithm \citep{MCS},
because sampled images cannot be fully deconvolved without
violating the sampling theorem.  Instead, the images are partially deconvolved,
meaning that we aim at an improved instead of an infinite resolution. As a
consequence, the PSF in the deconvolved image ("target" PSF) can be chosen in such
a way that it is well sampled. In practice, the shape of the target PSF is chosen
to be a circular Gaussian.  The description of the new algorithm is presented in
Cantale et al. (A\&A submitted). The new features, which specifically benefited from
the deconvolution of the EDisCS data, are an efficient regularization
strategy, a new minimization, and a proper noise treatment. The new
regularization uses a multi-scale approach to eliminate the high frequencies where
no signal is present. Special care was also taken in the proper noise treatment. An incorrect estimate of the noise may lead to significant errors
in the final photometry.

The algorithm uses super-resolved PSFs, constructed from several stars in the
field of view (two to six stars, depending on the cluster), with a smaller pixel scale
than in the original image. The stars were selected from the brightest stars,
without neighbours and close to the cluster centre, in such as way that the PSF
corresponds to the region in which the galaxies are located.  We chose the pixel
size to be four times smaller than the original size and a target PSF sampled with two
of these pixels at FWHM. In other words, the resolution in our deconvolved images
is 0.1\arcsec\ with a pixel size of 0.05\arcsec, which is similar to the
data from the ACS at the HST. All our images are therefore fully numerical (as opposed to described by an
analytical function), and the only prior on the light distribution imposed by the
regularization is a pixel-to-pixel continuity.

Figure ~\ref{Demo} shows a few examples of deconvolution over the full redshift
range of our sample. The galaxies were chosen to span a broad range of
angular sizes, compactness, and number of substructures.  We show the
$I$-band images of the FORS2 at the VLT, their deconvolved counterparts, the residual maps, and the
corresponding F814W images obtained with the ACS at the HST. The residual maps are the difference between
the original VLT image and its deconvolution, in units of photometric noise. These
are displayed between $-3\sigma$ (white) and $+3\sigma$ (black). The red circles
correspond to the aperture applied to remove the bulge light, as explained in
Sect.~\ref{aperphot}. The complexity of the galaxy morphologies and the small-size
features such as star-forming regions, faint spiral arms, or even dust lanes are
unveiled by the deconvolution procedure. They are all confirmed by the comparison
with the ACS images. Most of the time, the residual maps reveal only negligible
structures, which illustrates the excellent quality of our procedure. Only the regions
with very peaked luminosity profiles are difficult to fit because
of the
regularization and yield high residuals.

\begin{figure}
\hskip -7pt
\includegraphics[width=9.2cm]{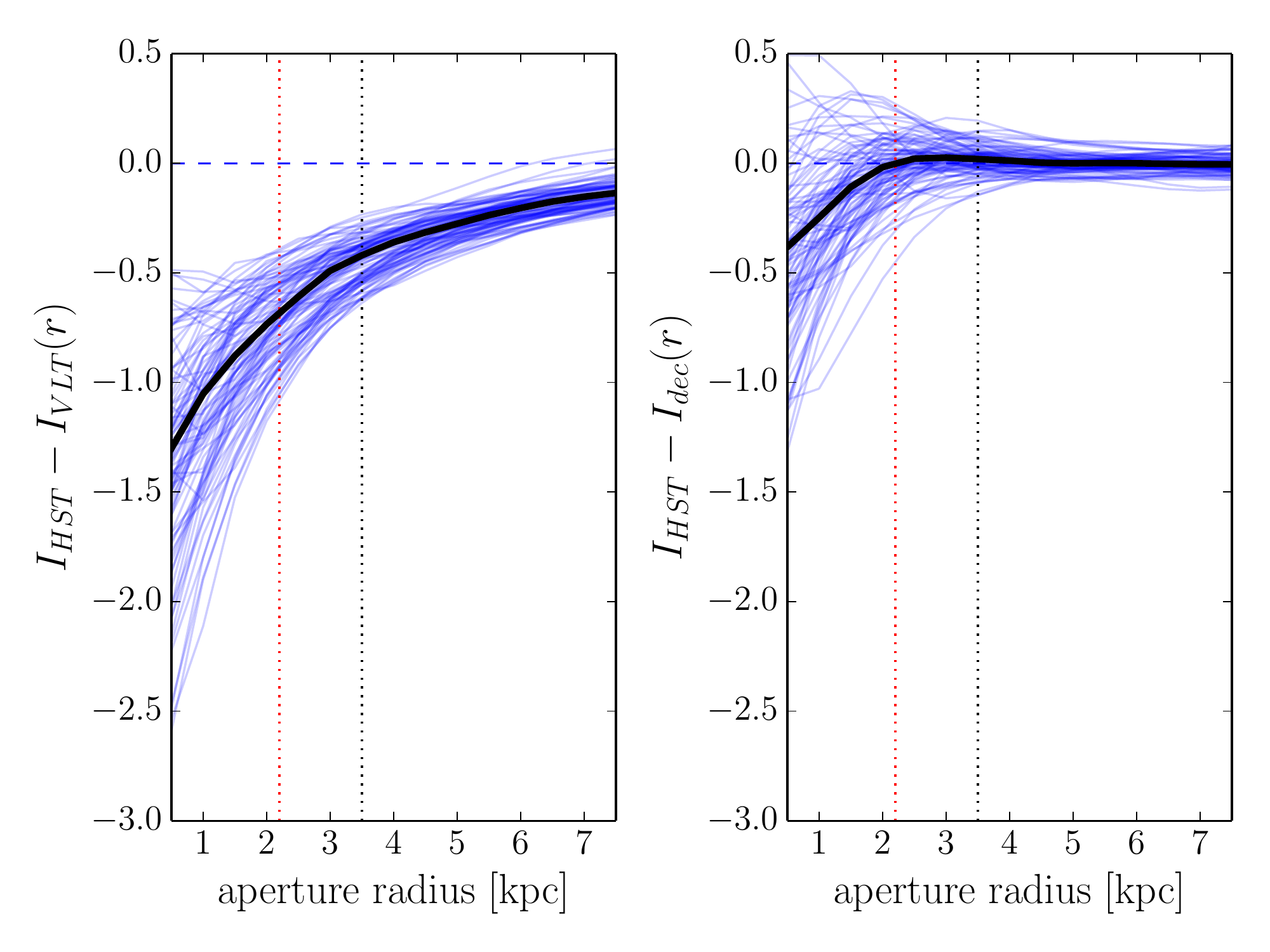}
\caption{{\it Left:} Differences between the aperture photometry performed on images obtained with the ACS at the HST and on the \textup{\textup{{\it \textup{original}} }}FORS2 images. {\it Right:} The difference between the aperture photometry performed on the
  {\it \textup{deconvolved}} FORS2 and the images of the ACS
at the HST.  Ninety percent of the ACS PSF energy is
  contained inside a 2.2 kpc radius, shown as the dotted red line.  The measurements made on the HST and the VLT-deconvolved images
  down to 3.5 kpc from the centre of the galaxy (black dotted line) agree excellently well.
}
\label{dejong_prof}
\end{figure}

\subsection{Aperture photometry}
\label{aperphot}

We performed aperture photometry with {\tt SExtractor} \citep{Bertin1996}.  The
centroids and the shapes of the global apertures were drawn from the original
$I$-band FORS2 data, taken as reference images, the photometric measurements were
subsequently carried out on the $V$-band and $I$-band deconvolved frames at a
resolution of 0.1\arcsec. The apertures used to compute the total object
magnitudes are the Kron-like elliptical aperture from {\tt SExtractor}.The
photometric error bars are given by the photon noise in the aperture:
\begin{equation}
m_{\rm err} = 2.5\cdot \mathrm{log}_{10}\left(1+ \frac{\sqrt{F_{\rm aper}+\sigma_{\rm sky}^2}}{F_{\rm aper}}\right),
\end{equation}
where $F_{\rm aper}$ is the total photon number counts in the aperture, that is, the
counts corresponding to the sum of the sub-exposure images and converted into
electrons; $\sigma_{\rm sky}$ is the sky noise in the full image, corresponding to
the standard deviation in the empty regions of the original image, as estimated
on the full image. This is the same as the image used during the deconvolution.

The flux of the disc component was recovered by subtracting the bulge contribution
from the galaxy total luminosity. The total galaxy flux was measured in an
elliptical aperture, while the flux of the galaxy bulge was calculated in a circular
aperture. The apertures were identical in $V$ and $I$ bands.

This method obviously requires the size of bulge aperture to be carefully chosen.
It must be sufficiently small to include most of the disc component, but large
enough to account for the diversity of bulges along the Hubble sequence. To fix this minimum size, we used the great variety of bulge and disc profiles of the
sample of face-on nearby spirals reported in \cite{Jong1996b} to build model galaxies with
redshifts, signal-to-noise ratios, angular sizes, and resolution similar to those
in our sample.  By varying the aperture radii, we calculated the flux of all simulated
galaxies. There is no PSF at this stage. We requested a maximum 5\% error on the
recovered disc fluxes, that is, a negligible pollution by the bulge. This corresponds to 0.05
mag, which is smaller than our photon noise errors (see below), and an aperture of
$r = 1.5$~kpc. Then, we evaluated the impact of the PSF. Although the
deconvolution brings the images to a much improved spatial resolution, there is
still a PSF to be taken into account, since we deconvolve to a finite
resolution. This means that the deconvolved data represent the true light
distribution, convolved by the Gaussian PSF of 0.1\arcsec FWHM. This, combined
with the necessary regularization inherent to the deconvolution, which also causes
spreading of the light, prevents the deconvolved data from reaching the true light
distribution in the inner parts of very sharp profiles. This affects the
photometry of the bulges, with a leakage of their light beyond the physical
theoretical limit. The left panel of Fig.~\ref{dejong_prof} presents the
difference between the $I$-band magnitudes integrated in the images obtained with ACS at the HST and in the
original FORS2 images as a function of the aperture radius, while the right panel
of the same figure presents the difference between the aperture magnitudes
calculated in the images of the ACS at the HST and in the final deconvolved $I$-band FORS2
images.  As revealed by the large difference between the HST and original-FORS2
magnitudes, the bulge light leaks onto the disc from the inner regions to far
out. The deconvolution
does not deal easily with sharp unresolved profiles. This is a consequence of our
choice not to introduce any a priori analytical profile: unresolved features, such
as point sources or bulges, conflict with the regularization. Nevertheless,
after deconvolution, the HST and the VLT (deconvolved) magnitudes agree on
average to within $\sim$0.05~mag (rms) in an aperture as small as 0.3\arcsec. This
corresponds to a physical aperture of 2.2~kpc, which we represent by a dotted red
line for the highest redshift galaxies.  We note that we cannot
conclude for objects below this limit because the HST PSF itself affects the profiles: within a 0.3\arcsec\ aperture, the
encircled energy of the HST PSF is about 90\%, which corresponds to a 0.1 mag.

Taking the still large scatter between the HST and deconvolved-VLT
magnitudes at $r = 0.3$\arcsec\ into account, we conservatively set the bulge aperture limit to
$r=0.5$\arcsec (the angular sizes apply to our redshift range).  This ensures that
the contamination of the disc by the bulge light is smaller than 0.05 mag. This
remains also true for the $V$-band images because bulges tend to be smaller in the
blue. We wish to compare galaxies at different redshifts, therefore
we carried out our
measurements on physical instead of on angular scales, that is, $r_{\rm
  ap,in}=3.5$~kpc. The corresponding angular apertures (red circles in
Fig.~\ref{Demo}) were calculated for each galaxy based on its redshift. They were typically between
0.45\arcsec\ ($z=0.8$) and 0.6\arcsec\ ($z=0.4$). The galaxy sizes extend
from 0.5\arcsec\ to 2.5\arcsec\, with a mean at 1\arcsec.  For the smallest galaxies, only 25-30\% of their surface was therefore
kept
to retrieve the photometry of the discs. However,
even though the uncertainty became large, the measurement was still
possible. Photometric errors were tracked down at all steps of the analysis.

\begin{figure}
\hskip -7pt
\includegraphics[width=9.2cm]{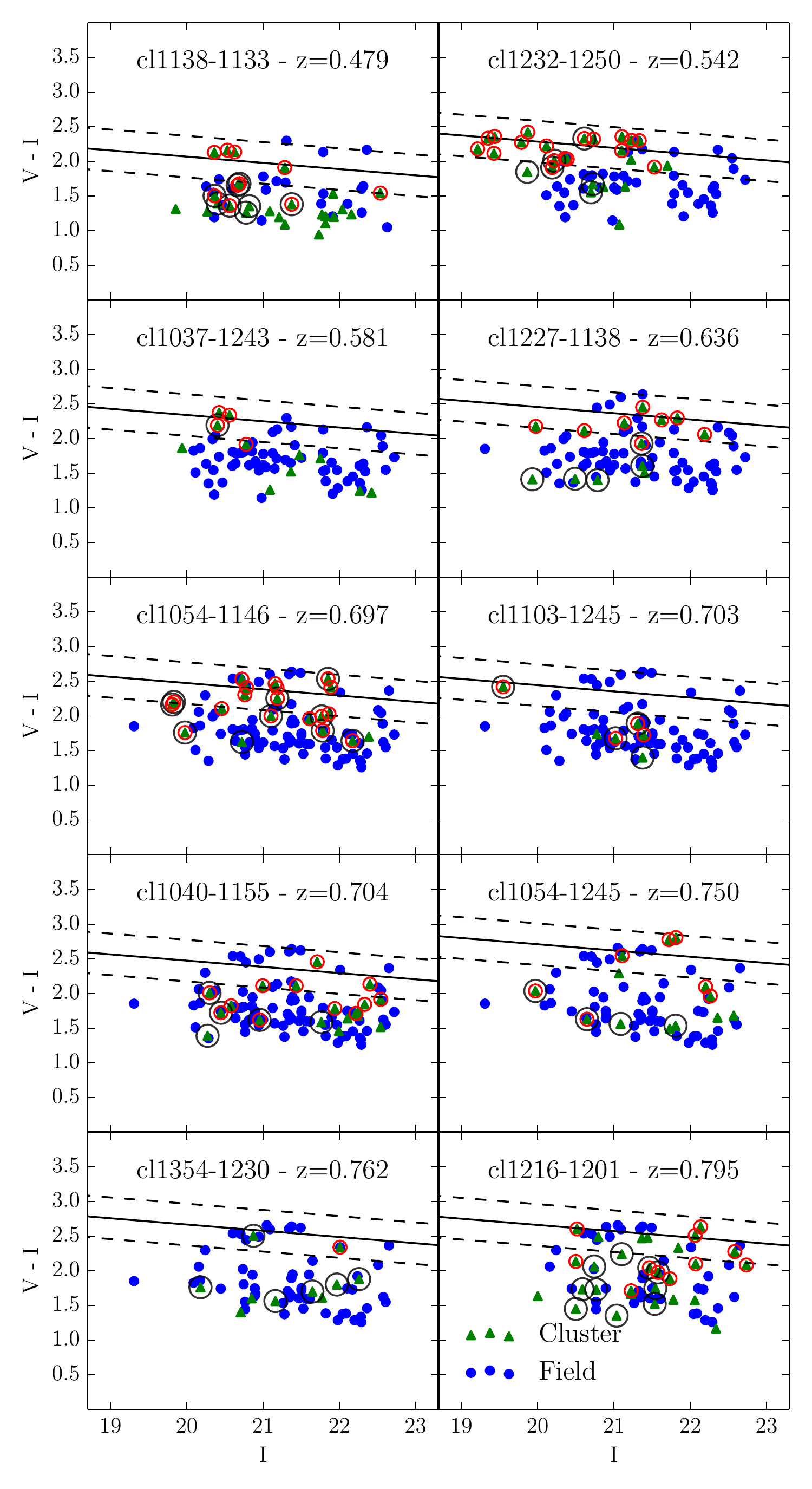}
\caption{Colour-magnitude diagrams of the spiral galaxies in the ten EDisCS
  clusters. In each panel, the green triangles indicate the cluster
  members. The blue points are the field galaxies within $\pm 0.03$ the cluster
  redshift with or without HST imaging (see Sect.~\ref{data}). The solid thick
  lines represent the red sequence best-fit relations, using all the cluster
  members and a bi-weight estimator \citep{Beers1990}, and assuming a fixed slope
  of $\alpha=-0.09$ as in \citet{DeLucia2007}. The dashed lines correspond to a
  $\pm0.3$ mag deviation from this best-fit line. The red circles identify
  galaxies with discs redder than their field counterparts, i.e., at same redshift
  and with same morphological type, as identified in Fig.~\ref{fit_all}. The black
  circles show the galaxies that were detected at 24~$\mu$m with {\it Spitzer}.}
\label{cmd}
\end{figure}

\section{Colours of the discs}
\label{results}

Figure ~\ref{cmd} shows the colour-magnitude diagrams of the spiral galaxies in
the ten \ediscs\ clusters.  At this stage, colours and magnitudes are still from
the original EDisCS photometry, meaning that they are integrated over the full
galaxies before deconvolution. In each panel, we identify the cluster members and
field galaxies with redshift $z = z_{\rm cl} \pm 0.03$. This field galaxy
selection was chosen to avoid crowding in the diagram. In each panel, we show the
position of the red sequence of the clusters calculated as in \cite{DeLucia2007}.
Field and cluster spirals cover very comparable locii in the CMDs, whereby we
exclude strong biases in the comparison of the properties of two populations.

\begin{figure}[t!]
\hskip -5pt
\includegraphics[width=9.3cm]{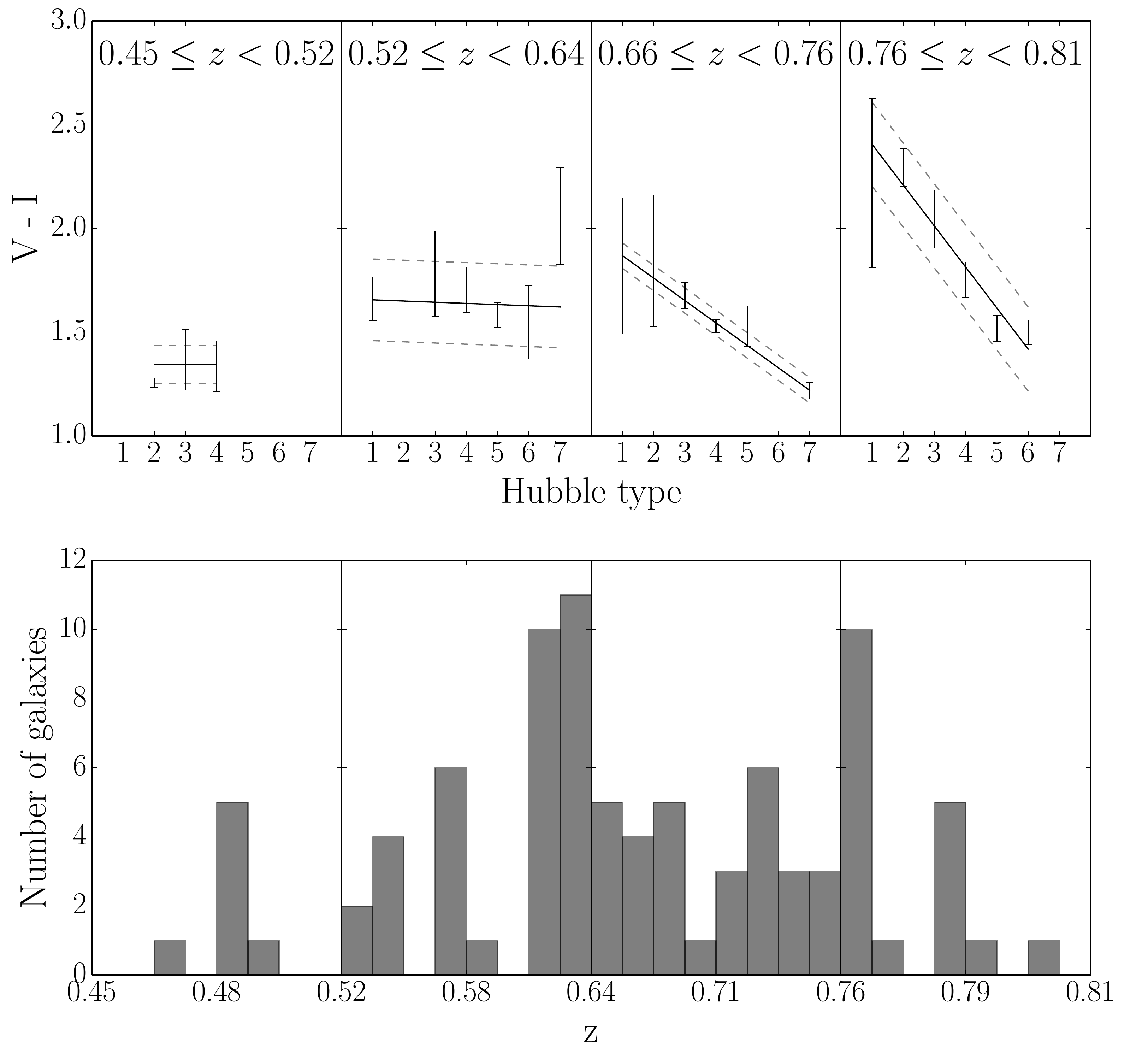}
\caption{Colour-morphology relation for the discs of field galaxies at different redshifts.
In each of the upper panels the solid lines show the best fit of the relation that is
obtained by keeping both the slope and the $y$-intercept as free parameters. The
dashed lines show the 1-$\sigma$ confidence interval, computed from the covariance
matrix of the fits. These fits are used to infer the relative colour properties of
field and cluster discs. The lower panel shows the redshift distribution of the
galaxies inside each redshift bin.}
\label{fit}
\end{figure}

\begin{figure}[h]
\hskip -7pt
\includegraphics[width=9.0cm]{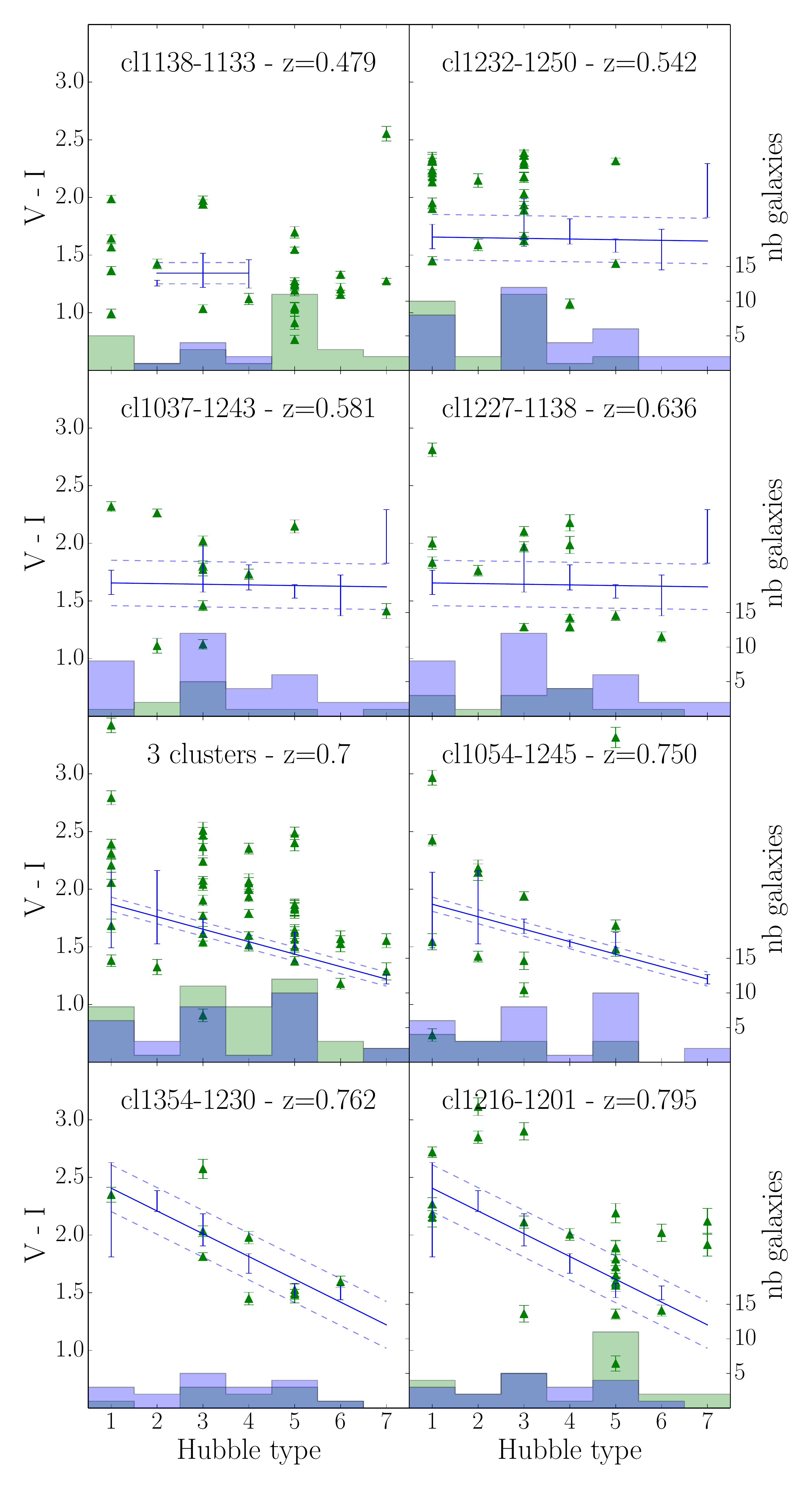}
\caption{Colours of the cluster spiral discs, shown with green triangles, as a
  function of the galaxy Hubble types. The fits of the relations obtained for the
  field galaxies from Fig.~\ref{fit} are shown in blue. The histograms show the
  number of galaxies at each Hubble type for the field (blue) and the clusters (green). The scale is on the right-hand vertical axis.  }
\label{fit_all}
\end{figure}

\subsection{Comparison between field and cluster population}

We now explore the relation between the colours of the\textup{ {\it \textup{discs}}} and the
morphological types of their parent galaxies in the field and
clusters. This time colours are obtained after and by the deconvolution.

Figure~\ref{fit} presents the mean colour of the spiral discs in the field and its
dispersion as a function of the galaxy Hubble type. The error bars correspond to
the dispersion around the mean value at each morphological type.  In the rare
cases of only one galaxy, we plot its photometric error.

We grouped the galaxies per redshift bin by considering the position of the $V$ and
$I$ filters with respect to the 4000\AA\ break.  Although the $V$ and $I$
  filters probe gradually bluer parts of the galaxy spectra from the highest to
  the lowest redshift, $V-I$ nevertheless  brackets the 4000\AA\ break
  over the full redshift range. The resulting redshift partition yields four
groups, $0.45 \le z<0.52$, $0.52\le z<0.64$, $0.66\le z < 0.76$, and $0.76\le
z<0.81$ and 89 field galaxies. The field comparison at $0.45\leq z < 0.52$ 
is only used for cl1138.2-1133. In each of the redshift groups, galaxies
  have masses between $\sim$$10^{9}$ and 10$^{11}$ M$_{\odot}$ as calculated with
  iSEDfit \citep{Moustakas2011}. 

Discs become clearly bluer towards later morphological types at z $\gtrsim$0.64 when
the $V$ filter gathers a majority of the light at a restframe wavelength shorter than
3500\AA.   The slope of the relation between $V-I$ and the galaxy Hubble type
becomes steeper with increasing redshift, which is essentially due to the reddening of the discs of
the earliest Hubble type galaxies. This is expected from the shift in zero-point
of the red sequences \citep{DeLucia2007}, $+0.4$ mag from z$\sim$0.5 to z$\sim$0.8.
 
The relations between the colours of the discs and the galaxy Hubble types in the
field are indicated again in Fig.~\ref{fit_all}, where we present the distribution
in $V-I$ for the discs of spirals in clusters.  The three clusters cl1054.4-1146
($z=0.6965$, $\sigma_{\rm cl} = 589$ \kms), cl1040.7-1155 ($z=0.7020$,
$\sigma_{\rm cl} = 418$ \kms) and cl1103.7-1245b ($z=0.7029$, $\sigma_{\rm cl} =
242$ \kms) are grouped together in one single bin at $\langle z \rangle =0.7
$. We provide the number of galaxies per Hubble type at each redshift with
histograms.

Figure~\ref{fit_all} is used to evaluate whether the colours of spiral discs differ
between field and cluster environments. This is done by comparing -- at each
Hubble type -- the $V-I$ disc colours of the cluster galaxies to the
colour-morphology relations obtained for the field sample. We can then define
three categories: At a given Hubble type, a cluster galaxy disc is defined as
blue (red) when its $V-I$ colour, including its uncertainty, is bluer
(redder) by more than $1\sigma$ (error on the slope on the fit in the field) than
the mean value obtained in the field for this morphology. Cluster galaxies with
disc colours consistent with their field counterparts (same morphology, same
redshift) are called green.

\subsection{Spatial distribution}

\begin{figure}[h]
\hskip -7pt
\includegraphics[width=9.2cm]{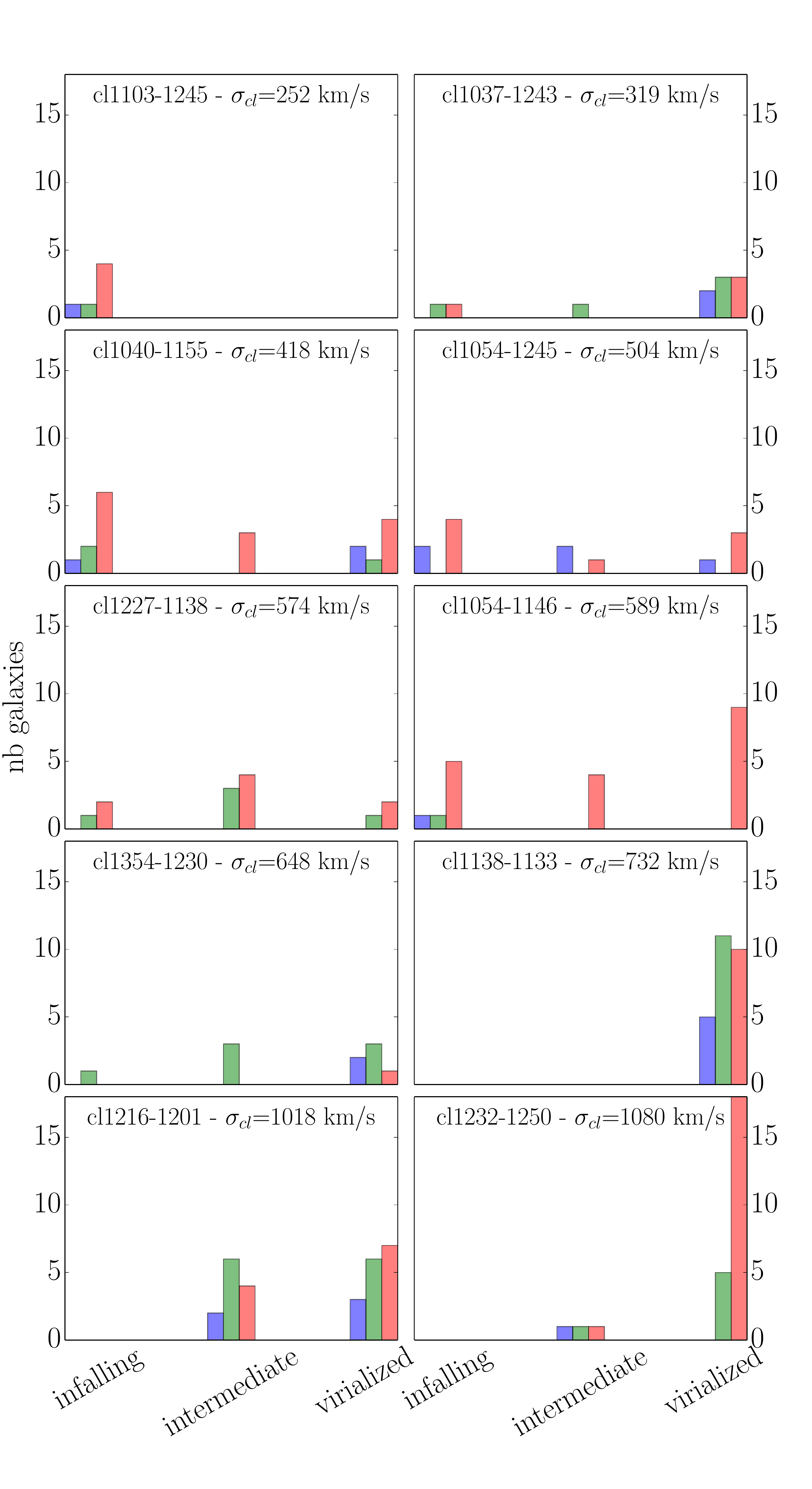}
\caption{Distribution of the colour of the spiral discs in clusters. The bar sizes
  represent the number of galaxies inside each galaxy category: infalling,
  intermediate, and virialized. Blue stands for cluster discs 1$\sigma$ bluer than
  their field counterparts. Red shows the number of cluster discs 1$\sigma$ redder
  than their field counterparts, and green illustrates the number of cluster discs
  with colours compatible with those of their field counterparts.}
\label{vir_colour_all}
\end{figure}

\citet{Mahajan2011} distinguished between virialized, infall, and backsplash
classes of galaxies based on their radial velocity and projected clustercentric
distances. We applied their scheme to our galaxy sample, using the probability of
a galaxy to be virialized, P$_{\rm{vir}}$ from their Table~2.  Virialized
galaxies have clustercentric distances smaller than the cluster virial radius,
infalling systems prefer large projected distances and high velocities, while
the backsplash particles are mostly found just around the virial radius and have
low absolute line-of-sight velocities.  In the following, we define three
categories: galaxies with P$_{\rm{vir}} > 0.7$ are defined as "virialized",
galaxies with P$_{\rm{vir}}\sim 0.5$ are "intermediate'' since they have as much
probability to be infalling as to be already virialized. The remaining galaxies
are ``infalling''.  Figure~\ref{vir_colour_all} summarizes how the colours of discs
in cluster compare to discs in the field for each of these three classes.  The
galaxy counts per cluster are presented as a function of the parent cluster
velocity dispersion.  Because we compare the disc colours to the mean field
population within 1$\sigma$, the signature of a cluster disc population compatible
with the field would be a normal distribution centred on the green bin (68\% of
the galaxies). This is not what is observed: Figure~\ref{vir_colour_all} reveals a
clear bias of the distribution of the cluster galaxies towards redder discs for
the majority of the clusters.

\begin{figure*}[h]
\centering
\includegraphics[width=16cm]{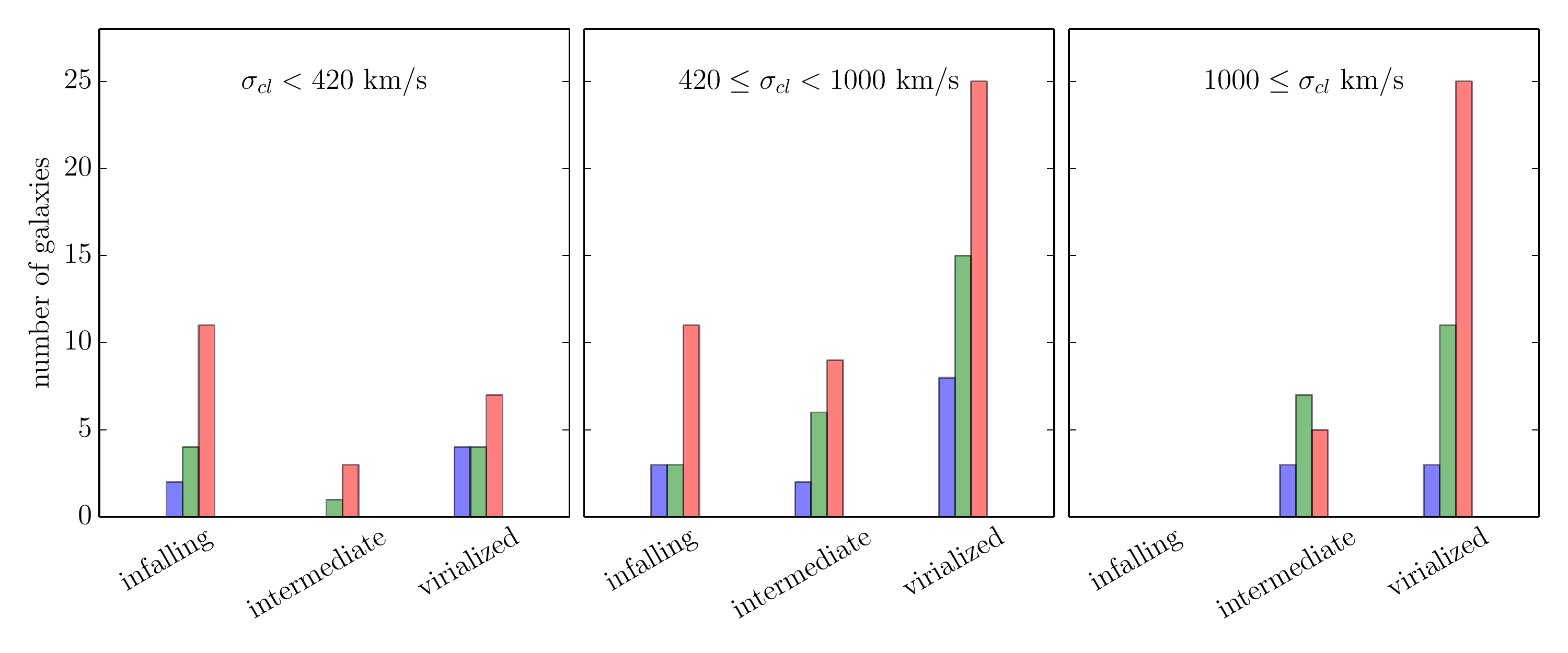}
\caption{Same as Fig.~\ref{vir_colour_all}, but with the clusters binned in velocity dispersion.}
\label{vir_colour}
\end{figure*}

\begin{figure*}[h]
\centering
\includegraphics[width=16cm]{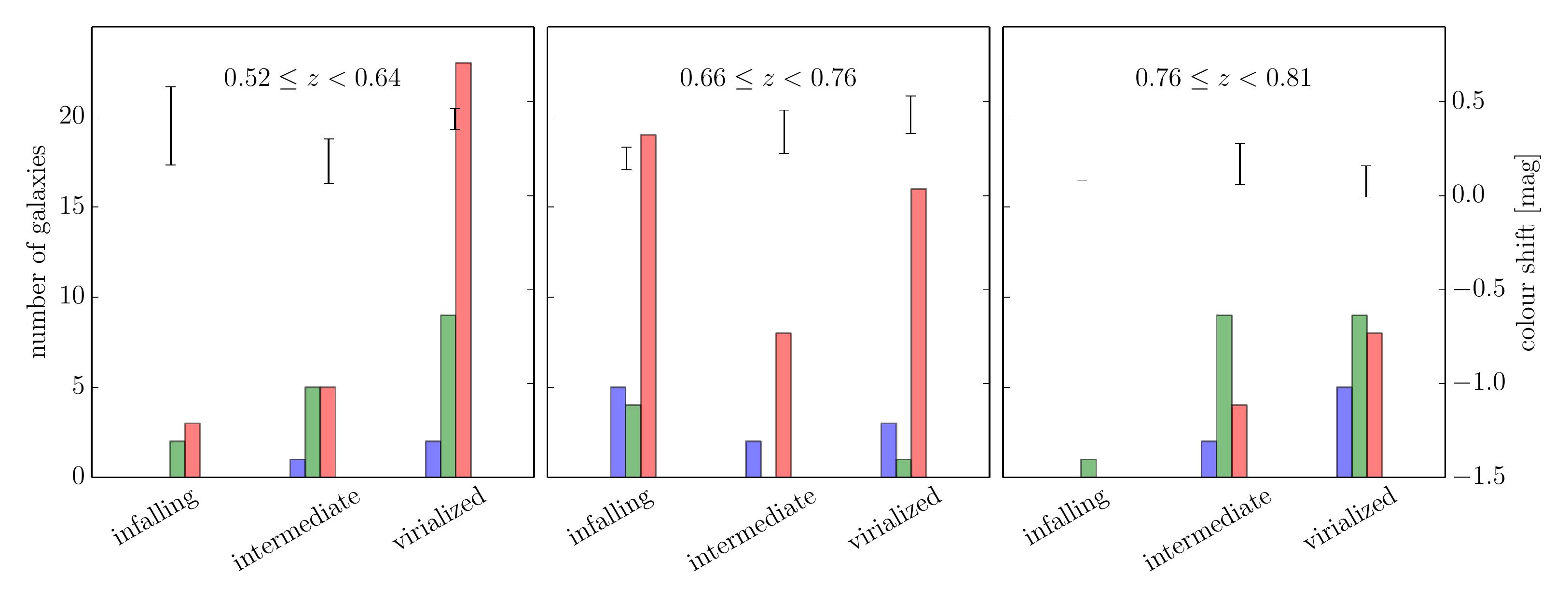}
\caption{Same as Figs.~\ref{vir_colour_all} and \ref{vir_colour}, but with the clusters
  binned in redshift. The mean colour shift, $\delta(V-I)$, between the field disc
  colours and the cluster disc colours is given for each galaxy type and for each
  redshift bin. The scale of these shifts is provided on the right-hand vertical axis. The error bars
  correspond to the dispersion around this mean value.}
\label{vir_colour_z}
\end{figure*}

\subsection{Cluster mass}

To increase the statistical significance of our result, we now bin the clusters in
velocity dispersion, $\sigma_{\rm cl}$, in Fig.~\ref{vir_colour}.  We split the
sample into three categories: low ($\sigma_{\rm cl} < 420$ \kms), moderate ($420
\le \sigma_{cl} < 730$ \kms), and high ($\sigma_{\rm cl} > 1000$ \kms) velocity
dispersions. The fractions of red discs are explicitly given in Table
\ref{tab:redfrac} for the full sample together with the infall, intermediate, and
virialized galaxies. The prominence of galaxies with red discs clearly depends
neither on the mass of their parent cluster nor on whether the galaxies are in the
infall, intemediate, or virialized group.

We also tested the robustness of our results by removing one cluster at a time
in each $\sigma_{\rm cl}$ bin and checking any change in the global features of
Fig.~\ref{vir_colour}. Cl1232.5-1250 ($z=0.54$) is
the only cluster that affects the distribution of its group. Given its size and relatively
low redshift, we essentially only sample  its virialized region.
It is indeed one of the two most massive clusters of our sample, the
other one being cl1216.8-1201 ($z=0.79$) at much higher redshift.  The
group of the $\sigma_{\rm cl} > 1000$km/s clusters obviously
suffers from degeneracy between
evolution with time and mass.

Without cl1232.5-1250 and cl1216.8-1201, that is, restricting the analysis to
structures with $\sigma_{\rm cl}$ from 252 km/s to 732 km/s, we find that
$\sim$50\% of the spiral galaxies have a disc redder than the mean field
population regardless of the cluster mass in which they reside.  This is much
more than the expected 16\% (fraction outside the 1-$\sigma$ probability
distribution) we expect to measure if the cluster and field  populations were
similar.

Cl1232.5-1250 clearly stands out because its fraction of spirals with red discs is
larger than the rest of the investigated structures. Similarly, in our hightest
redshift bin, the fraction of red discs is higher in cl1216.8-1201
($\sigma$ = 1080 km/s) than in cl1354.2-1230 ($\sigma$=648 km/s). This suggests that
that either extremely dense environments act as particularly efficient star
formation suppressors or that galaxies in these massive clusters have been
perturbed longer.

\subsection{Look-back time}

Figure~\ref{vir_colour_z} groups galaxies in redshift bins. For each cluster
galaxy we calculated the difference in colour, $\delta (V-I)$, between its colour
and the mean colour of the field galaxies in its redshift bin $\delta (V-I) =
[V-I]_{\rm cl} - [V-I]_{\rm field}$. The mean and the dispersion of these colour
shifts are indicated, with values given on the right y-axis of the figure. The
error bars are the standard deviation around the mean $\delta (V-I)$.

Figure~\ref{vir_colour_z} reveals a sharp increase in the number of reddened discs
below z = 0.76.  The lowest and highest z bins are dominated in number by
cl1232.5-1250 and cl1216.8-1201, respectively. Nevertheless, cl1216.8-1201 and
cl1354.2-1230 in the high-z bin share consistent disc properties in that they
encompass a smaller fraction of spirals with red discs than cluster at lower
redshift. Similarly, removing cl1232.5-1250 from the low-z bin diminishes the number
of red discs in its groups, but they still represent the dominant population.
In conclusion, the increase in the number of galaxies with reddened discs below
z=0.76 is a robust result of our analysis. Its general significance
must be assessed in more detail by conducting similar studies at higher redshift,
as our small number of clusters at z $\ge$ 0.75 makes us sensitive to possible
variance.

\subsection{Extinction}

The observed colour reddening of the cluster discs might be caused
in principle by
some highly inclined and therefore dust-extinct galaxies. We know from
Fig.~\ref{fig:population} that our cluster and field global populations follow the
same distribution of inclination angles. However, these distributions may vary
within each redshift, $\sigma_{\rm cl}$ or virialization bin and thereby cause a
shift in $V-I$.  To determine this potential bias, we used the inclination angles
computed on the images obtained with the ACS at the HST \citep{Simard2009} to exclude galaxies with angles
larger than 70$^\mathrm{o}$. This inclination is available for 158 cluster members
and 77 field galaxies. Once discarded, the highly inclined galaxies, 63 field
galaxies and 114 cluster galaxies remain and our results are unchanged.  In
conclusion, strongly inclined galaxies do not drive the observed reddening of some
of the discs in clusters.

\begin{table}[!h]
\caption{Fractions of cluster spiral galaxies with discs redder than their field   counterparts, $\mathrm{N_{red}}$, as a function of cluster velocity dispersion and     redshift. The parts (a) and (b) of the table correspond to Figs.~\ref{vir_colour} and \ref{vir_colour_z}, respectively.  $\mathrm{N_{tot}}$ is the total number of     galaxies. $\mathrm{N_{infall}}$ is the number of infalling galaxies,     $\mathrm{N_{inter}}$ the number of intermediate galaxies, and    $\mathrm{N_{vir}}$ the number of virialized galaxies.}
\label{tab:redfrac}
\renewcommand{\arraystretch}{1.25} 
\begin{tabular}{lccc}
(a) & & & \\ 
\hline
$\sigma_{cl}$ [\kms] & $\mathrm{N_{red}/N_{tot}}$ & $\mathrm{N_{red}/(N_{infall}+N_{inter})}$ & $\mathrm{N_{red}/N_{vir}}$\\
\hline
$\sigma_{cl}<420$        &  0.58 & 0.67 & 0.47\\
$420 \le \sigma_{cl}< 1000$ &  0.55 & 0.59 & 0.52\\
$1000 \ge \sigma_{cl}$      &  0.56 & 0.33 & 0.64\\
\hline
\end{tabular}
\\[12pt]
\begin{tabular}{lccc}
(b) & & & \\ 
\hline
$z$ & $\mathrm{N_{red}/N_{tot}}$ & $\mathrm{N_{red}/(N_{infall}+N_{inter})}$ & $\mathrm{N_{red}/N_{vir}}$\\
\hline
$0.52 \le z<0.64$ &  0.62 & 0.50 & 0.68\\
$0.66 \le z<0.76$ &  0.74 & 0.71 & 0.80\\
$0.76 \le z<0.81$ &  0.32 & 0.25 & 0.36\\
\hline
\end{tabular}
\vspace*{2pt}
\end{table}

\section{Discussion}
\label{discussion}

In all redshift groups, galaxies in clusters clearly deviate from a normal
distribution centred on the properties of the field galaxies. In the $0.52 \le
z<0.64$ and $0.66 \le z < 0.76$ groups, $\sim$ 60\% and 70\%\ of
discs, respectively, are redder than their field counterparts.  Above $z=0.76$, the discs in the
field and in clusters appear to have more comparable stellar populations, with a
total fraction of red discs of only $\sim$ 30\%. Calculating the binomial
probabilities of such events rejects at 99\% confidence level that the cluster
populations could be built from the field sample.

Figure \ref{cmd} reveals that this redder disc population is not restricted to the
cluster red sequences and can also be found in the cluster blue sequence. This is
particularly clear with increasing lookback time.  We flag the
galaxies in Fig. 4 that were detected by Spitzer at 24 $\mu$m \citep{Finn2010}, when those
observations were available. At $z=0.79$, our 80\% star formation rate
completeness limit is $\sim$ 14 M$_{\odot}$/yr, while at $z=0.48$ we reach 5 M$_{\odot}$/yr.

\begin{figure*}[h]
\centering
\includegraphics[width=1.0\textwidth]{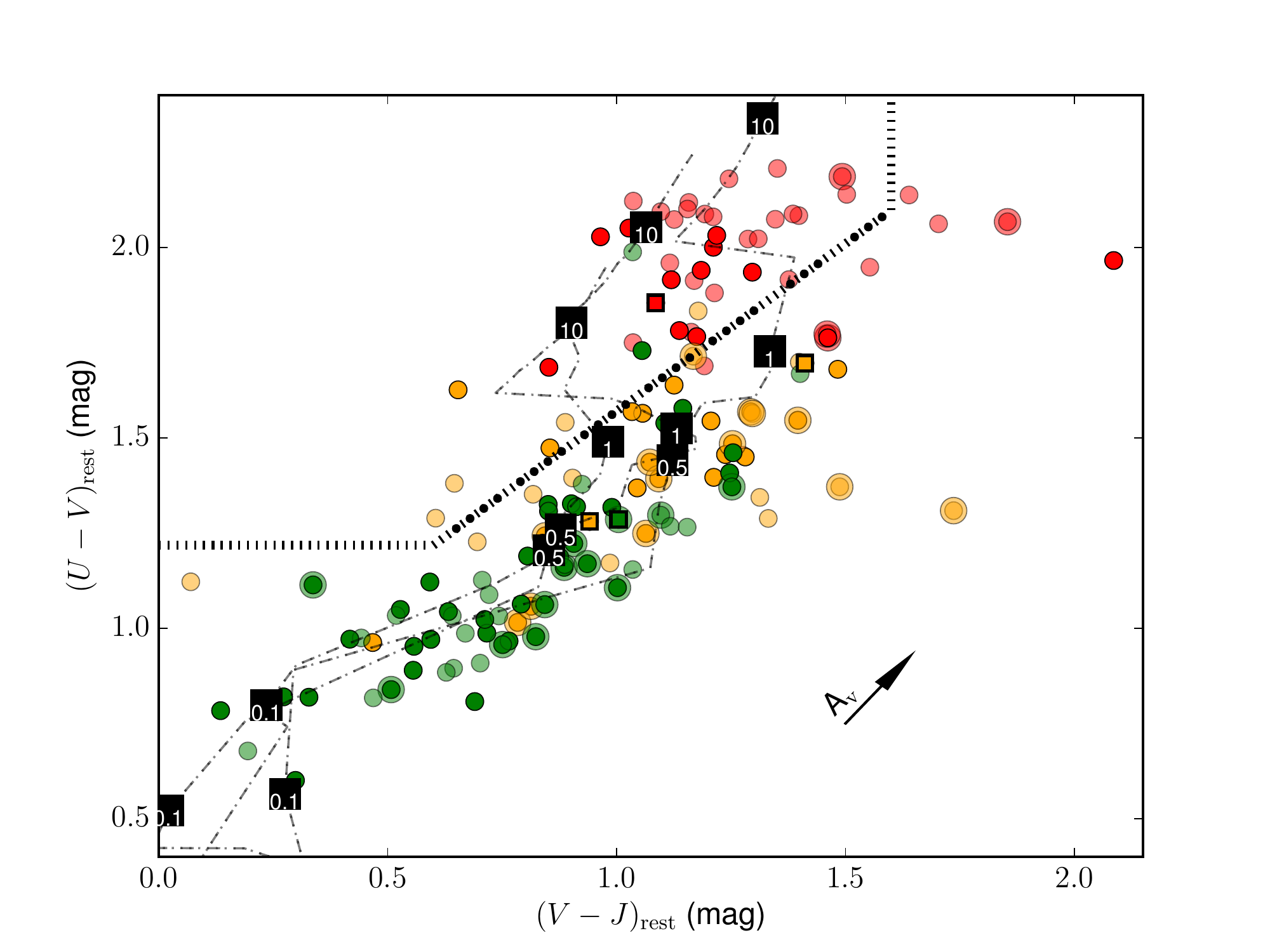}
\caption{Rest frame $U-V$ versus $V-J$ diagram. Galaxies with disc colours compatible with their field counterparts are shown in green.
Galaxies that are still forming stars, but whose discs are nonetheless redder than their field counterparts, 
are plotted in orange. Galaxies with red discs that are
on the red sequence are shown in red. Large circles identify galaxies that have been detected at 24$\mu$m with Spitzer and have logL$_{\mathrm {IR}}$ $\ge$ 10.91 (SFR $\sim$ 14 M$_{\odot}$/yr. Plain colours identify face-on galaxies. The three squares correspond to the three galaxies in Fig.\ref{fig:spectra}. The models of
\cite{maraston1998} and
\cite{maraston2003}  are shown as dot-dashed lines, and a few indicated ages for the single stellar population are indicated.}
\label{fig:UVJ}
\end{figure*}

Figure \ref{fig:UVJ} presents the distribution of our sample disc galaxies in the
two restframe colour diagram $U-V$ versus $V-J$. The $U$, $V$, and $J$ magnitudes
are total, meaning they are integrated over all galaxies and are derived from our ground-based
photometry \citep{White2005} using the technique of \cite{Rudnick2006} and
\cite{Rudnick2009}. The colours of the points distinguish between the cluster
galaxies i) that are located on the red sequence with discs redder than their field
counterparts (in red). These red-on-redseq galaxies (for red discs galaxies on the
red sequence) compose $\sim$24\% of our full sample. ii) Galaxies
that are not located on the red
sequence, but with discs redder than their field counterparts (in
orange). These red-not-on-redseq galaxies compose another $\sim$24\% of our sample
of cluster galaxies. Finally, iii) galaxies that are not located on the red sequence with disc colours
similar to their field counterparts (in green). Following \cite{williams2009}, we
indicate the dividing line between star-forming and passive galaxies.  It
is striking to see how well the properties of the discs reveal the properties of
their parent galaxies.

\subsection{Red-sequence galaxies}

\begin{figure*}
\includegraphics[width=2.6in]{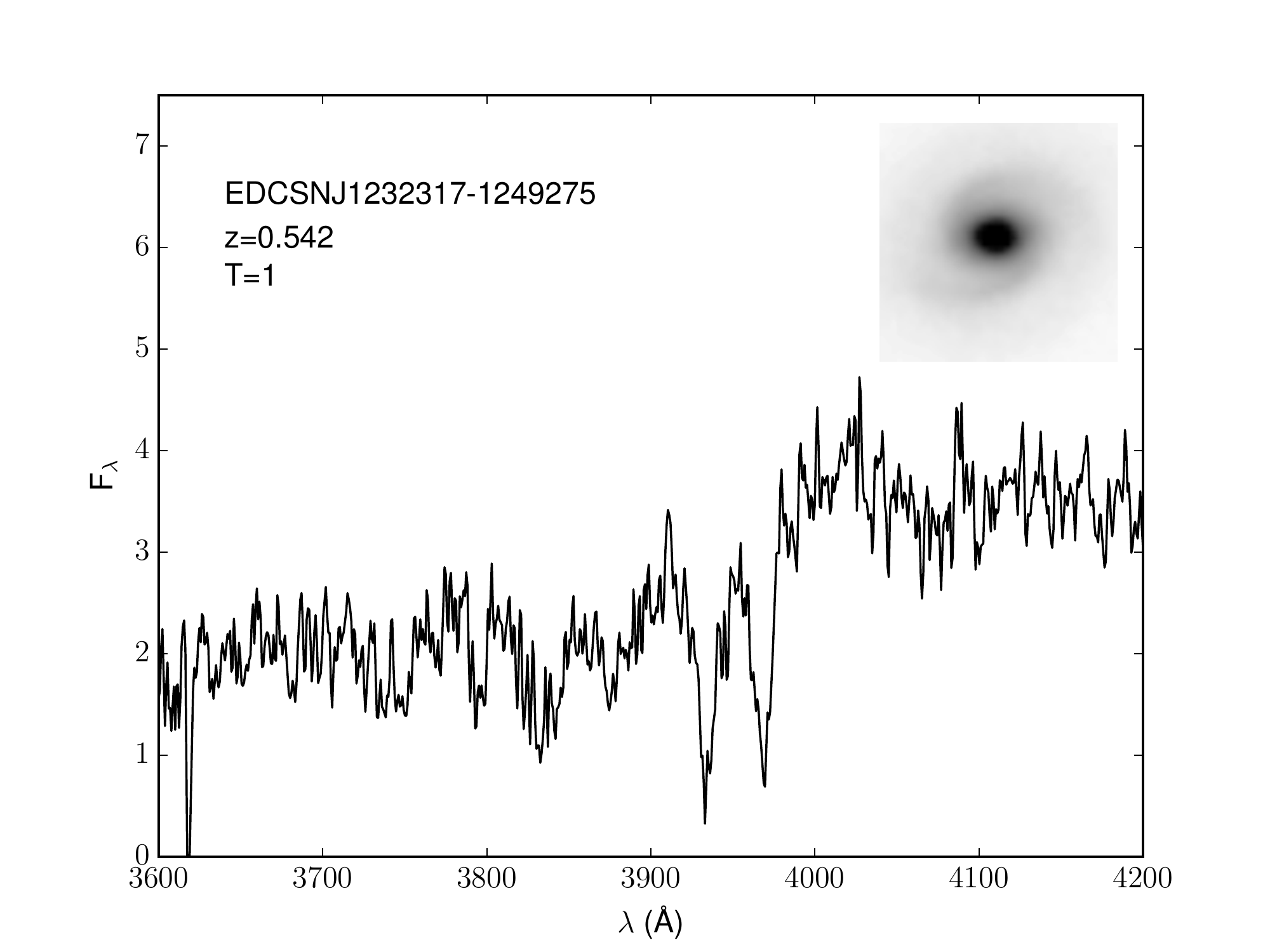}\hspace{-2.em}%
\includegraphics[width=2.6in]{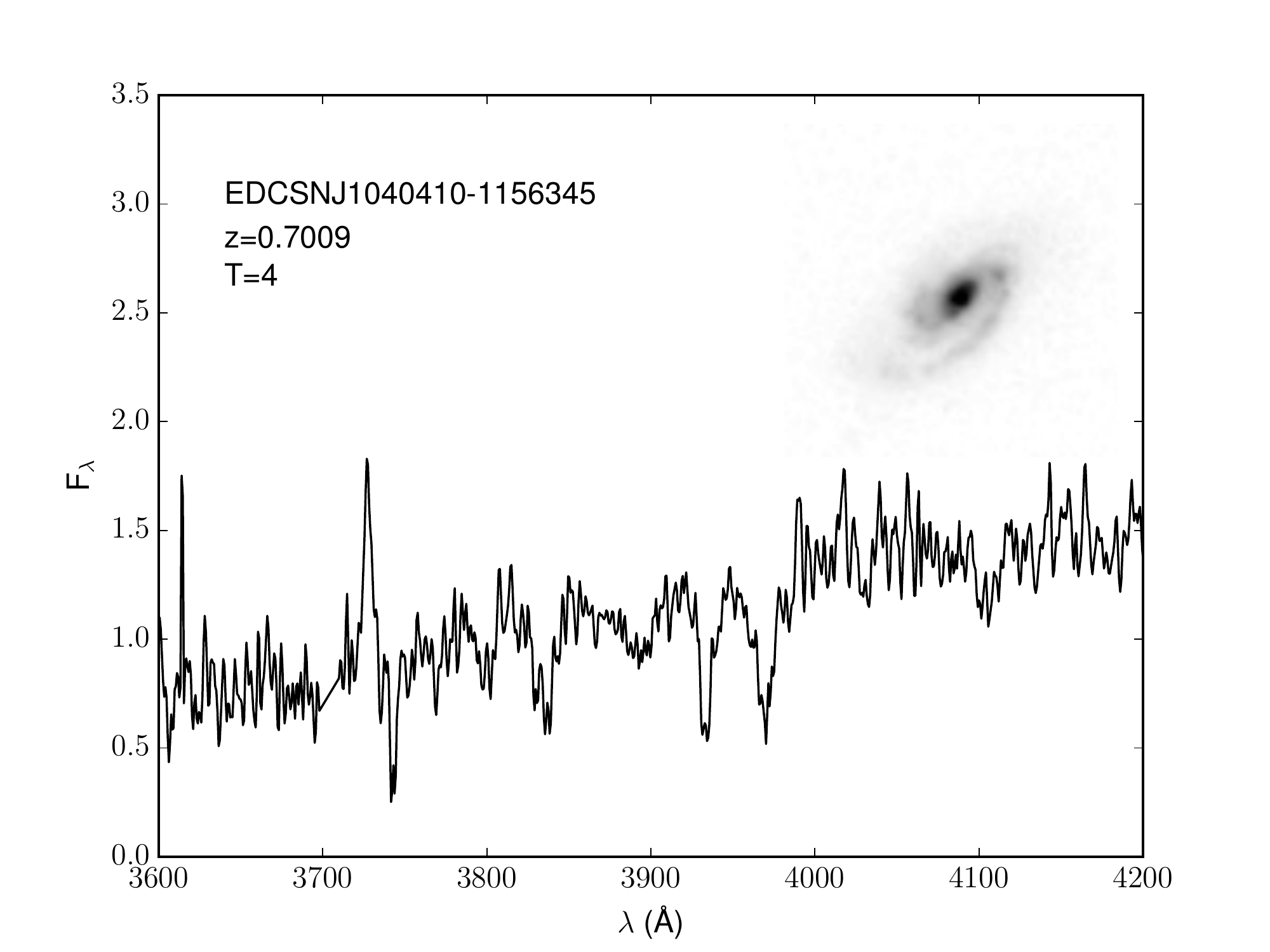}\hspace{-2.em}%
\includegraphics[width=2.6in]{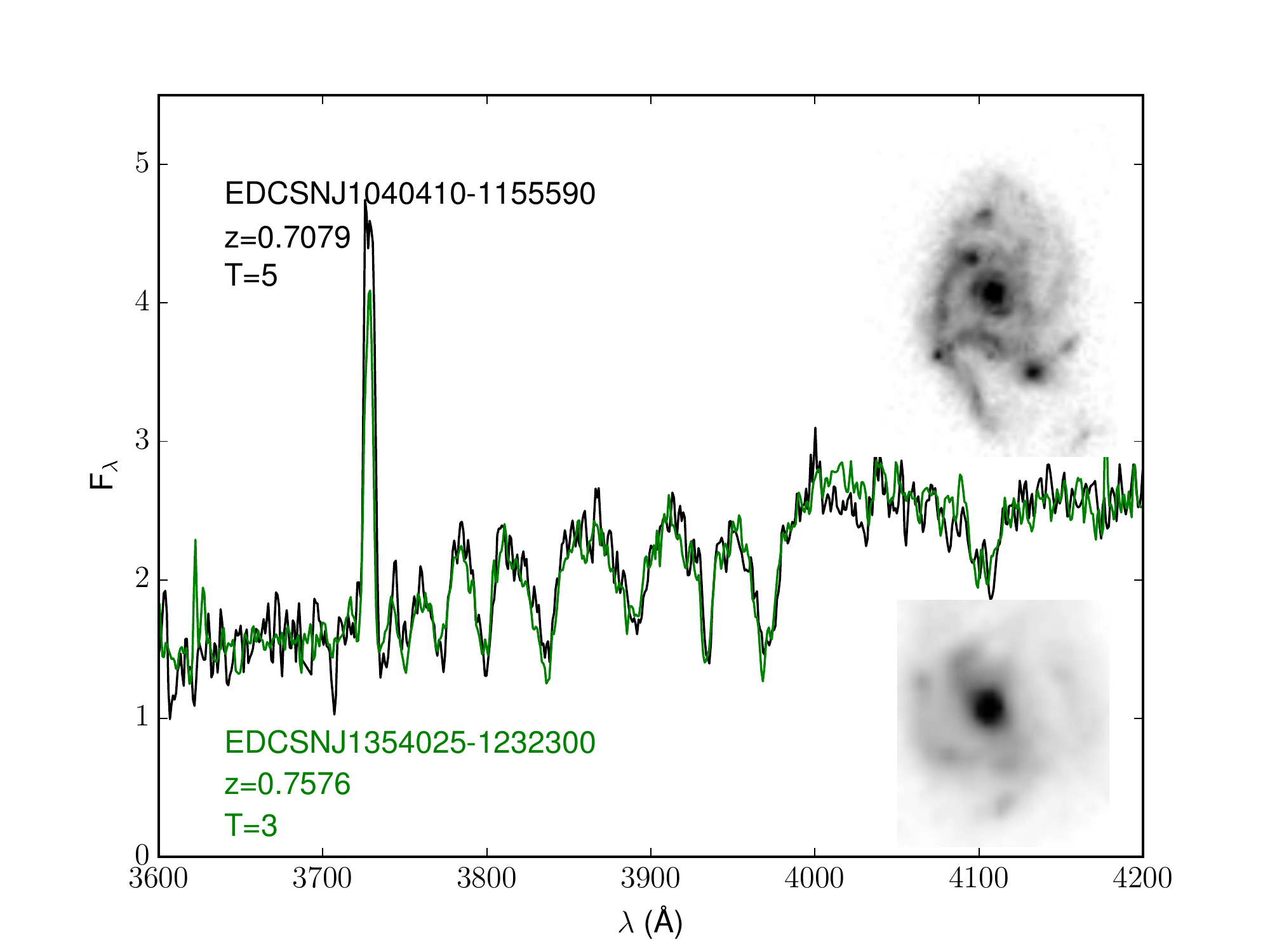}
\caption{Three examples of FORS2 spectra. The name of the galaxy, redshift, and
  T-type are indicated. The left panel displays a representative spectrum of a
  galaxy on the red sequence, whose disc is redder than its field counterparts. The
  middle panel shows the spectrum of a galaxy in the blue sequence, but forming stars at a low rate and with a red disc. The right panel displays in black the
  spectrum of a galaxy that actively forms stars, but with redder
disc   than its field counterparts. For comparison we show in green a galaxy with a
  field-like disc that exhibits the same spectral feature and colours as the
  galaxy in black, but with an earlier morphological T-type.}
\label{fig:spectra}
\end{figure*}

The vast majority (80\%) of the galaxies that lie on the cluster red sequences are
also located in the region of the ($U-V$, $V-J$) plane where the systems that
have stopped forming stars are expected to be. Visual inspection of the images of
the remaining galaxies reveal that these 20\% are composed of inclined
dusty star-forming galaxies. Most of them are detected by Spitzer. If
our sensitivity at 24$\mu$m had been higher, we would probably have detected them
all.  The left panel of Fig. \ref{fig:spectra} provides a typical example of a red
spiral galaxy whose star formation has been quenched, shows its HST/F814W image, and its
FORS2 spectrum. While the spiral arms can still be identified,  dust lane
and clumps of young stars are absent. The optical spectrum is typical of an old
($>$ 5 Gyr) stellar population with strong CN molecular bands below
3900$\AA$.  Interestingly, this type of galaxies is found in all clusters
regardless of their mass. Some of them belong to the category of infalling
galaxies, but most belong to the virialized groups
defined in the previous section. This means that galaxies can be pre-processed before they are
trapped in the cluster potential well. Correcting for the galaxies that lie on
the red sequence because of their inclination and high dust content, the fraction
of genuinely passive spiral galaxies constitutes 20\% of our sample. This result
combines ten clusters and corresponds to the fractions of \cite{Moran2006}, 27\%,
in the cluster Cl0024+ 17 at z=0.4 (possibly including galaxies with residual star
formation), 17\% in \citep{Wolf2009} in A901/2 at z=0.17. Red passive galaxies are
also observed in low-density environments \citep{Poggianti1999, Goto2003,
  Masters2010}, but at lower frequencies.

\subsection{Blue-sequence galaxies}

The right panel of Fig. \ref{fig:spectra} shows a typical example of
red-not-on-redseq galaxy with red disc, but whose colour and magnitude do not place
it on its cluster red sequence. This time, dusty star-forming regions appear
clearly, although not on all arms. The FORS2 spectrum reveals high-order Balmer
lines typical of stellar population younger than 5 Gyr.

We indicate in Fig. \ref{fig:UVJ} the evolutionary tracks of star clusters in the
($U-V$, $V-J$) plane as provided by the models of \cite{maraston1998} and
\cite{maraston2003}. We carefully calibrated the zeropoints of these models
against the observations of globular clusters of \cite{Frogel1980} and
\cite{Bica1986a}. We show three metallicities [Z/Z$_{\odot}$] $-0.33$, 0., and
0.35, which are representative of the range of mean metallicities of spiral
galaxies.  Stellar clusters are -- to first-order approximation -- single stellar
populations with fixed age and metallicity, and they are meant to be
representative of the mean light-weighted properties of the galaxies. At this stage, we do not
need a more detailed population synthesis that takes the
full metallicty and age range of the stellar population into
account \citep[e.g.][]{jablonka1990}.

The key lesson of this exercise is that while extinction runs along the
sequence of star-forming galaxies and might naturally be responsible for some of
the scatter in their distribution, the aging of the stellar population follows the same
direction. This holds until some time between 1 and 5 Gyr, when the stellar
population colour-colour evolution becomes perpendicular to the star-forming
galaxy sequence, towards bluer $V-J$ and redder $U-V$.

Undoubtely, red-not-on-redseq galaxies are forming stars, hence the question arises whether
their discs are redder by dust or because their star formation rates have
decreased compared to their field counterparts, which increases the proportion of ageing
stars.

\begin{figure}[h]
\centering
\includegraphics[width=9.0cm]{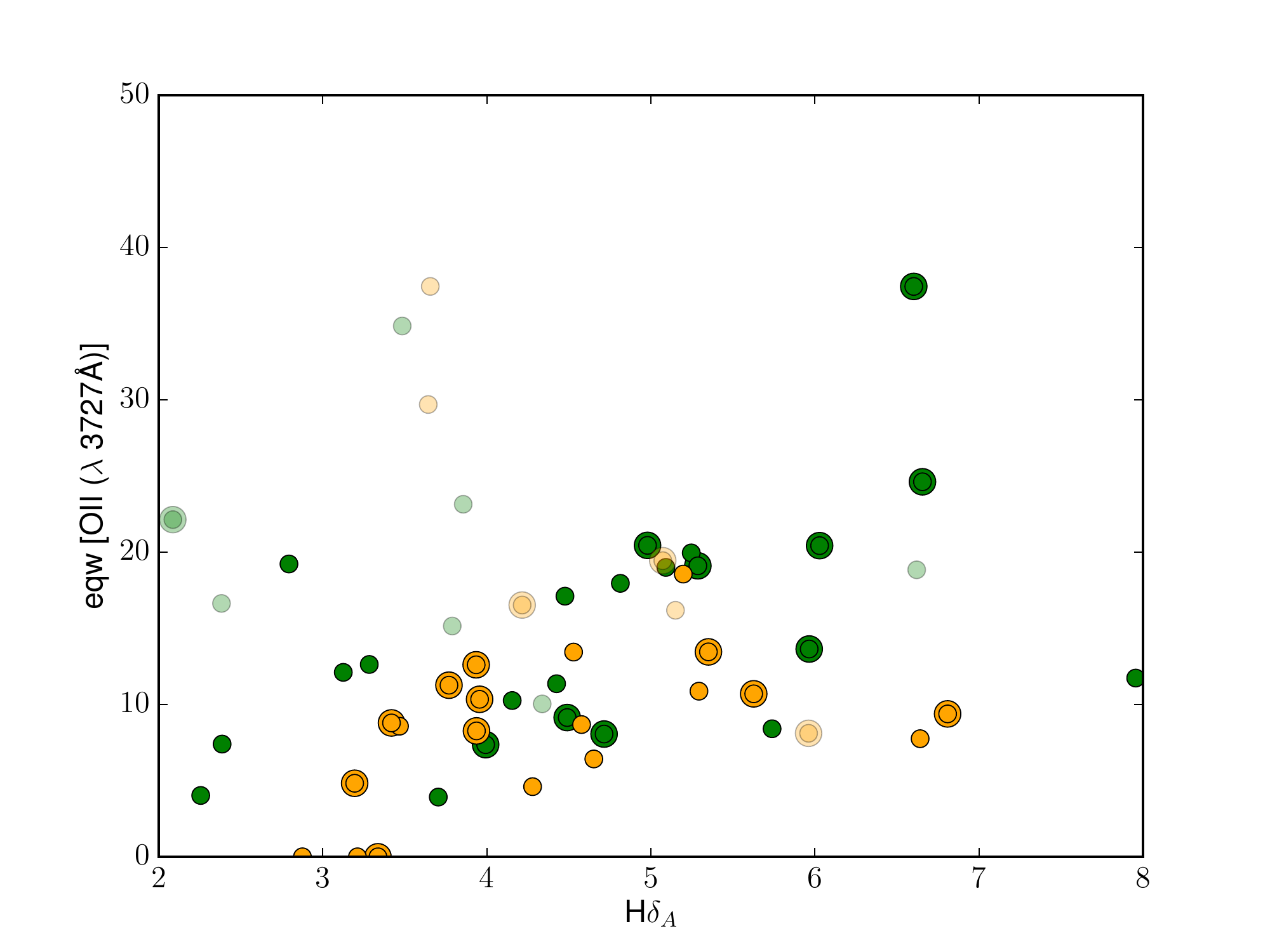}
\caption{OII equivalent width versus H$\delta_A$ diagram. Galaxies with disc colours compatible with their field counterparts are shown in green.
Galaxies that are still forming stars but whose discs are redder than their field counterparts are plotted in orange.}
\label{fig:OIIHdA}
\end{figure}

To answer this question, we first  consider the [OII] $\lambda 3727 \AA$ emission
line and compare its equivalent width in red-not-on-redseq galaxies and in cluster
galaxies that have discs comparable to field ones. The result is shown in
Fig. \ref{fig:OIIHdA}, where we show the distribution of the [OII] equivalent width
as a function of the H$\delta_A$ equivalent width \citep{Balogh1999, Worthey1997}. We show
all galaxies, but restrict our analysis to galaxies with low inclination angles
($<$70 degrees), which are displayed in plain colours, while the others are shown
in transparency, to avoid spurious effects to differences in the orientation of
the slits with respect to the galaxy major axis.  The distribution of the
red-not-on-redseq galaxies in the OII-H$\delta_A$ plane is similar to the galaxies
compatible with their field counterparts, but lacks the high [OII]
equivalent widths at high H$\delta_A$.

We indicate with large circles galaxies that have been detected by Spitzer at
24$\mu$m, with logL$_{\mathrm {IR}}$ $\ge$ 10.91 (SFR $\sim$ 14 M$_{\odot}$/yr, the
80\% completeness limit of the highest redshift of our cluster, cl1216.8-1201
\citep{Finn2010}. For galaxies with stellar masses above 10$^{10}$ M$_{\odot}$ and
with a redshift above z=0.6 for which we have sufficient statistics, 59\% of the
field-like spirals have SFR $\sim$ 13 M$_{\odot}$/yr as measured in the far-infrared
against 62\% for the red-not-on-redseq galaxies.  Restricting
the sample to the exact same
limit in OII equivalent width (7\AA $\le$ eqw(OII) $\le$ 15\AA) for the two sets
of galaxy populations results in detections of 50\% and 57\%, respectively. This
means that if dust enshrouds star formation, it does it in a similar way in both
types of galaxies. Because dust is genuinely associated with star
formation, it can add dispersion in the distribution of galaxies along the star-forming galaxy sequence in the $U-V$ vs $V-J$ plane, but it is not by itself at
the origin of the mean location of the red-not-on-redseq and field-compatible
galaxies.

There are essentially two degrees of evolution among the red-not-on-redseq
galaxies: i) Those that are still actively forming stars, which would
hence not be detected in a survey focussing on star formation rate indicators alone as
galaxies on the way to be quenched. However, their activity already
is or has been
for a period diminished with respect to their morphology. A concrete example is
provided by EDCSNJ1040410-1155590 (red-not-on-redseq, 0.7079,
log(M$_{\star}$)=10.94), whose spectrum is shown in the right panel of
Fig.~\ref{fig:spectra}. It has the same $U-V$ ($\sim$1.25) and $V-J$ ($\sim$1.)
colours as EDCSNJ1354025-1232300 (field-like, z=0.7576,
log(M$_{\star}$)=11.08). These galaxies are identified with squares with black
borders in Fig.~\ref{fig:UVJ}. Both galaxies also have very similar OII and
H$\delta_A$ equivalent widths, and their spectra are superimposed remarkably
well over each other in Fig.~\ref{fig:spectra}. EDCSNJ1040410-1155590 was visually
classified as T-type=5 (Sc), while EDCSNJ1354025-1232300 has T-type=3 (Sb).
EDCSNJ1040410-1155590 is therefore considered red for its Hubble type. ii) The other
type of red-not-on-redseq galaxies is an accentuated version of the previous one,
in which galaxies have only residual star formation, such as EDCSNJ1040410-1156345
($U-V$ $\sim$1.4 and $V-J$ $\sim$ 1.6) in the middle panel of
Fig.~\ref{fig:spectra}. Their spectral signature is a mix between CN absorption
and residuals of the high-order Balmer lines, revealing an ageing stellar
population. As such, these red-not-on-redseq galaxies are the closest systems to
being fully quenched.

Thirty percent of the red-not-on-redseq galaxies are in the infalling group
of galaxies, adding evidence that a fraction of galaxies are pre-processed in
external structures before reaching the cluster cores.  

It is important to stress that while the spatial coverage of each cluster varies
simply because a given FORS2 field of view covers larger clustercentric distances of
a low-mass cluster than a massive one, the category of the virialized galaxies
essentially encompasses only galaxies within $0.5\times \mathrm{R}_{200}$ of all
structures, hence allowing a proper comparison between all groups and clusters. This
category is the most populated one. Farther out, our statistics is admittedly
poorer. Nevertheless, there is a fair indication that a significant fraction of the spirals in
small structures ($\sigma_{\rm cl} < 420$ \kms, 2 groups at $z=0.70$, one at
$z=0.58$) have been perturbed before they reach the cores of the group.

The red-on-redseq galaxies have preferentially early-type morphologies, while the
red-not-on-redseq galaxies have later (T-Type$\ge$5) ones. This is easily
understandable since there is a genuine strong correlation between morphologies,
colours, and star formation rate \citep[e.g. Fig. 3 in][]{Kennicutt1998,
  Jansen2000}. Once their SFR are lower, the earlier types may soon reach the red
sequence, while the more strongly star-forming systems still remain on the blue
sequence or the green valley of the colour-magnitude diagrams.

\subsection{Timescales}

The time it takes for a galaxy to stop forming stars is a long-standing
question. Figure \ref{fit_all} shows that there is a range in intensity of disc
reddening that extends from a minor colour shift $\delta$$V-I$ to $\sim$0.5 mag in each
redshift bin.  This reflects the diversity of galaxy morphologies and associated
star formation history to which the variety of infall trajectories
onto the cluster central regions must be added.

It is helpful to form an idea of the ages of the stellar populations we consider. The Maraston models plotted in Fig. \ref{fig:UVJ} gives to first
approximation light-weighted mean ages of our sample of cluster spirals between
$5\times$10$^8$ to 1$\times$10$^9$yr when galaxies are located on the star-forming system sequence and above 5$\times$10$^9$yr otherwise. Since these
restframe colours are integrated over all galaxies, while colour differences
between field and cluster are calculated for the discs alone, we confirmed these
estimates by reproducing the $V-I$ colours of the discs alone, as shown in
Fig. \ref{fit_all}. To do this, we used the observed spectra of globular clusters
with metallicities from $-1.5$ to solar and open clusters with ages spanning
10$^8$ to 5$\times$10$^9$yr, from the database of \cite{Bica1986a,Bica1986b}, which
we redshifted.

It is interesting to note that 58\% of the galaxies in the virialized groups of
galaxies are still forming stars.  Some at a slower pace than others and already
redder than their field counterparts, but all active. Not only do these results
robustly confirm that galaxies are able to continue forming stars for some
significant period after being accreted into clusters, they also suggest that star
formation can decline on seemingly long timescales, as revealed by the existence
of red-not-on-redseq galaxies. These galaxies are still forming stars, but for
their colour to become redder, some of them must have diminished star formation
rates for several Gyr (1 to 5 Gyr) (e.g. Fig. \ref{fig:UVJ}).  The proportion of
younger and older stars must slowly change to favour the latter. Our conclusion
agrees with the findings of some recent works that favoured long (1-4 Gyr)
quenching timescales \citep{Balogh2000, Finn2008, deLucia2012, Taranu2014,
  Haines2015}.

\citet{Koopmann2004} analysed galaxies with reduced total star
formation in the Virgo cluster in which they took into account the Hubble type of
their galaxies as we do. They found that these systems have truncation (more
centrally concentrated H$\alpha$ extent than normal systems) rather than anemia
(low H$\alpha$ surface brightness across the disk), which causes the reduced total
star formation rates.  It is too early to conclude on this part for our sample,
although \cite{Jaffe2011}  found OII truncated
at 1-2$\sigma$ significance in the EDisCS spirals for a subsample of the present sample. Further investigations both in
OII and H$\alpha$ are planned.

\section{Conclusion}
\label{conclusion}

We presented the first analysis of the $V-I$ disc colours in cluster and field spiral
galaxies at intermediate redshift that is free of any prior on the shape of the
galaxy luminosity profiles. 

We deconvolved the $V$ and $I$  images of 172 cluster and 96 field spiral
galaxies obtained with FORS2 at the VLT with initial resolution of 0.48\arcsec\ to 0.85\arcsec\ and achieved a
final spatial resolution of 0.1\arcsec\ with 0.05\arcsec\ pixels;
this is close to the resolution of the ACS at the HST.  After removing the central 3.5 kpc of each galaxy to
avoid contamination by the bulges, we measured the $V-I$ colours of the discs and
compared them to the mean colour of the discs in field spirals. This comparison
was conducted per Hubble type and took both the photometric errors
and the dispersion of the relations obtained in the field into
account. A cluster galaxy disc
was considered to be red (blue) when its $V-I$ colour, including its
uncertainty, was redder (bluer) by more than $1\sigma$ (error on the slope on the
fit in the field) than the mean value obtained in the field for this morphology.

\begin{itemize}

\item We find a large portion of spiral galaxies in clusters with $V-I$ colours
  redder than those in the field at the same morphological type. The prominence of galaxies
  with red discs depends neither on the mass of their parent cluster nor on
  whether the galaxies are in the infall, intermediate, or already virialized
  regions.

\item In all redshift groups, galaxies in clusters have disc colours that clearly
  deviate from a normal distribution centred on the properties of the field
  galaxies. In the $0.52 \le z<0.64$ and $0.66 \le z < 0.76$ groups,
  $\sim$ 60\% and 70\%\ of discs, respectively, are redder than their field counterparts.  Above
  $z=0.76$, the discs in the field and in clusters appear to share more comparable
  stellar populations, with a total fraction of red discs of only $\sim$
  30\%. Calculating the binomial probabilities of such events rejects that the cluster populations could be built from the field
  sample at 99\%
  confidence level. Future studies in clusters at higher redshift are needed to assess the
  significance of the lower fraction of redder discs at $z > 0.76$ that we find.

\item Passive spiral galaxies constitute 20\% of our sample. They are located on the
  cluster red sequences. These systems are not abnormally dusty; they are are made of
  old stars and are located on the cluster red sequences. 

\item Another 24\% of our cluster galaxy sample is composed of galaxies that are
  still active and star forming, but have discs redder than their field
  counterparts. These galaxies are naturally located in the blue sequence of their
  parent cluster colour-magnitude diagrams.

\item The reddest of the discs in clusters must have stopped forming stars more
  than $\sim$ 5 Gyr ago. Some of them are found among infalling galaxies,
  suggesting that pre-processing is at work long before galaxies reach the cluster
  cores.

\item Our results confirm that galaxies are able to continue forming stars for
  some significant period of time after being accreted into clusters. They also
  suggest that star formation can decline on seemingly long timescales. Galaxies that are still forming stars but have discs redder than their field
  counterparts must have diminished their star formation rates for 1 Gyr to 5 Gyr
  to have redder colours.

\end{itemize}

With ongoing and future wide field surveys such as KIDS, DES, and Euclid, multi-band
photometry with matched spatial resolution will become of increasing
importance. This is done so far with profile fitting of individual bands,
for example with GIM2D \cite{Simard2009}, GALFIT \citet{Peng2011}, or by simultaneously
fitting many bands \citep[Megamorph,][]{Megamorph}. Direct deconvolution of the
data followed by aperture photometry offers an interesting alternative to model
fitting, as it does not depend on any prior on the shape of the galaxies to be
studied. Any complex structure is restored at high spatial resolution,
independently of any underlying model. Our deconvolution method is fast enough to
be applied in an automated way to large samples of galaxies. The work presented in
this paper may therefore be extended to much larger areas with limited or no HST
imaging, but where multi-band images are available from the ground.

\begin{acknowledgements}
This work was supported by the Swiss National Science Foundation (SNSF).
We warmly thank Yara Jaffe for timely and informative discussions.
\end{acknowledgements}

\bibliographystyle{aa}
\bibliography{main.bib}

\begin{thebibliography}{63}
\expandafter\ifx\csname natexlab\endcsname\relax\def\natexlab#1{#1}\fi

\bibitem[{{Aguerri} {et~al.}(2004){Aguerri}, {Iglesias-Paramo}, {Vilchez}, \&
  {Mu{\~n}oz-Tu{\~n}{\'o}n}}]{Aguerri2004}
{Aguerri}, J.~A.~L., {Iglesias-Paramo}, J., {Vilchez}, J.~M., \&
  {Mu{\~n}oz-Tu{\~n}{\'o}n}, C. 2004, \aj, 127, 1344

\bibitem[{{Bah{\'e}} \& {McCarthy}(2015)}]{Bahe2015}
{Bah{\'e}}, Y.~M. \& {McCarthy}, I.~G. 2015, \mnras, 447, 969

\bibitem[{{Balogh} {et~al.}(1999){Balogh}, {Morris}, {Yee}, {Carlberg}, \&
  {Ellingson}}]{Balogh1999}
{Balogh}, M.~L., {Morris}, S.~L., {Yee}, H.~K.~C., {Carlberg}, R.~G., \&
  {Ellingson}, E. 1999, \apj, 527, 54

\bibitem[{{Balogh} {et~al.}(2000){Balogh}, {Navarro}, \& {Morris}}]{Balogh2000}
{Balogh}, M.~L., {Navarro}, J.~F., \& {Morris}, S.~L. 2000, \apj, 540, 113

\bibitem[{{Bamford} {et~al.}(2007){Bamford}, {Milvang-Jensen}, \&
  {Arag{\'o}n-Salamanca}}]{Bamford2007}
{Bamford}, S.~P., {Milvang-Jensen}, B., \& {Arag{\'o}n-Salamanca}, A. 2007,
  \mnras, 378, L6

\bibitem[{{Beers} {et~al.}(1990){Beers}, {Flynn}, \& {Gebhardt}}]{Beers1990}
{Beers}, T.~C., {Flynn}, K., \& {Gebhardt}, K. 1990, \aj, 100, 32

\bibitem[{{Bertin} \& {Arnouts}(1996)}]{Bertin1996}
{Bertin}, E. \& {Arnouts}, S. 1996, \aaps, 117, 393

\bibitem[{{Bica} \& {Alloin}(1986{\natexlab{a}})}]{Bica1986a}
{Bica}, E. \& {Alloin}, D. 1986{\natexlab{a}}, \aap, 162, 21

\bibitem[{{Bica} \& {Alloin}(1986{\natexlab{b}})}]{Bica1986b}
{Bica}, E. \& {Alloin}, D. 1986{\natexlab{b}}, \aaps, 66, 171

\bibitem[{{Blanton} \& {Moustakas}(2009)}]{Blanton2009}
{Blanton}, M.~R. \& {Moustakas}, J. 2009, \araa, 47, 159

\bibitem[{{Christlein} \& {Zabludoff}(2004)}]{Christlein2004}
{Christlein}, D. \& {Zabludoff}, A.~I. 2004, \apj, 616, 192

\bibitem[{{Chung} {et~al.}(2009){Chung}, {van Gorkom}, {Kenney}, {Crowl}, \&
  {Vollmer}}]{Chung2009}
{Chung}, A., {van Gorkom}, J.~H., {Kenney}, J.~D.~P., {Crowl}, H., \&
  {Vollmer}, B. 2009, \aj, 138, 1741

\bibitem[{{de Jong}(1996)}]{Jong1996b}
{de Jong}, R.~S. 1996, \aap, 313, 45

\bibitem[{{De Lucia} {et~al.}(2007){De Lucia}, {Poggianti},
  {Arag{\'o}n-Salamanca}, {White}, {Zaritsky}, {Clowe}, {Halliday}, {Jablonka},
  {von der Linden}, {Milvang-Jensen}, {Pell{\'o}}, {Rudnick}, {Saglia}, \&
  {Simard}}]{DeLucia2007}
{De Lucia}, G., {Poggianti}, B.~M., {Arag{\'o}n-Salamanca}, A., {et~al.} 2007,
  \mnras, 374, 809

\bibitem[{{De Lucia} {et~al.}(2012){De Lucia}, {Weinmann}, {Poggianti},
  {Arag{\'o}n-Salamanca}, \& {Zaritsky}}]{deLucia2012}
{De Lucia}, G., {Weinmann}, S., {Poggianti}, B.~M., {Arag{\'o}n-Salamanca}, A.,
  \& {Zaritsky}, D. 2012, \mnras, 423, 1277

\bibitem[{{Desai} {et~al.}(2007){Desai}, {Dalcanton}, {Arag{\'o}n-Salamanca},
  {Jablonka}, {Poggianti}, {Gogarten}, {Simard}, {Milvang-Jensen}, {Rudnick},
  {Zaritsky}, {Clowe}, {Halliday}, {Pell{\'o}}, {Saglia}, \&
  {White}}]{Desai2007}
{Desai}, V., {Dalcanton}, J.~J., {Arag{\'o}n-Salamanca}, A., {et~al.} 2007,
  \apj, 660, 1151

\bibitem[{{Ebeling} {et~al.}(2014){Ebeling}, {Stephenson}, \&
  {Edge}}]{Ebeling2014}
{Ebeling}, H., {Stephenson}, L.~N., \& {Edge}, A.~C. 2014, \apjl, 781, L40

\bibitem[{{Finn} {et~al.}(2008){Finn}, {Balogh}, {Zaritsky}, {Miller}, \&
  {Nichol}}]{Finn2008}
{Finn}, R.~A., {Balogh}, M.~L., {Zaritsky}, D., {Miller}, C.~J., \& {Nichol},
  R.~C. 2008, \apj, 679, 279

\bibitem[{{Finn} {et~al.}(2010){Finn}, {Desai}, {Rudnick}, {Poggianti}, {Bell},
  {Hinz}, {Jablonka}, {Milvang-Jensen}, {Moustakas}, {Rines}, \&
  {Zaritsky}}]{Finn2010}
{Finn}, R.~A., {Desai}, V., {Rudnick}, G., {et~al.} 2010, \apj, 720, 87

\bibitem[{{Frogel} {et~al.}(1980){Frogel}, {Persson}, \& {Cohen}}]{Frogel1980}
{Frogel}, J.~A., {Persson}, S.~E., \& {Cohen}, J.~G. 1980, \apj, 240, 785

\bibitem[{{Fumagalli} \& {Gavazzi}(2008)}]{Fumagalli2008}
{Fumagalli}, M. \& {Gavazzi}, G. 2008, \aap, 490, 571

\bibitem[{{Gonzalez} {et~al.}(2001){Gonzalez}, {Zaritsky}, {Dalcanton}, \&
  {Nelson}}]{Gonzalez2001}
{Gonzalez}, A.~H., {Zaritsky}, D., {Dalcanton}, J.~J., \& {Nelson}, A. 2001,
  \apjs, 137, 117

\bibitem[{{Goto} {et~al.}(2003){Goto}, {Okamura}, {Sekiguchi}, {Bernardi},
  {Brinkmann}, {G{\'o}mez}, {Harvanek}, {Kleinman}, {Krzesinski}, {Long},
  {Loveday}, {Miller}, {Neilsen}, {Newman}, {Nitta}, {Sheth}, {Snedden}, \&
  {Yamauchi}}]{Goto2003}
{Goto}, T., {Okamura}, S., {Sekiguchi}, M., {et~al.} 2003, \pasj, 55, 757

\bibitem[{{Gray} {et~al.}(2009){Gray}, {Wolf}, {Barden}, {Peng},
  {H{\"a}u{\ss}ler}, {Bell}, {McIntosh}, {Guo}, {Caldwell}, {Bacon}, {Balogh},
  {Barazza}, {B{\"o}hm}, {Heymans}, {Jahnke}, {Jogee}, {van Kampen}, {Lane},
  {Meisenheimer}, {S{\'a}nchez}, {Taylor}, {Wisotzki}, {Zheng}, {Green},
  {Beswick}, {Saikia}, {Gilmour}, {Johnson}, \& {Papovich}}]{Gray2009}
{Gray}, M.~E., {Wolf}, C., {Barden}, M., {et~al.} 2009, \mnras, 393, 1275

\bibitem[{{Guti{\'e}rrez} {et~al.}(2004){Guti{\'e}rrez}, {Trujillo}, {Aguerri},
  {Graham}, \& {Caon}}]{Gutierrez2004}
{Guti{\'e}rrez}, C.~M., {Trujillo}, I., {Aguerri}, J.~A.~L., {Graham}, A.~W.,
  \& {Caon}, N. 2004, \apj, 602, 664

\bibitem[{{Haines} {et~al.}(2015){Haines}, {Pereira}, {Smith}, {Egami},
  {Babul}, {Finoguenov}, {Ziparo}, {McGee}, {Rawle}, {Okabe}, \&
  {Moran}}]{Haines2015}
{Haines}, C.~P., {Pereira}, M.~J., {Smith}, G.~P., {et~al.} 2015, \apj, 806,
  101

\bibitem[{{Halliday} {et~al.}(2004){Halliday}, {Milvang-Jensen}, {Poirier},
  {Poggianti}, {Jablonka}, {Arag{\'o}n-Salamanca}, {Saglia}, {De Lucia},
  {Pell{\'o}}, {Simard}, {Clowe}, {Rudnick}, {Dalcanton}, {White}, \&
  {Zaritsky}}]{Halliday2004}
{Halliday}, C., {Milvang-Jensen}, B., {Poirier}, S., {et~al.} 2004, \aap, 427,
  397

\bibitem[{{H{\"a}u{\ss}ler} {et~al.}(2013){H{\"a}u{\ss}ler}, {Bamford}, {Vika},
  {Rojas}, {Barden}, {Kelvin}, {Alpaslan}, {Robotham}, {Driver}, {Baldry},
  {Brough}, {Hopkins}, {Liske}, {Nichol}, {Popescu}, \& {Tuffs}}]{Megamorph}
{H{\"a}u{\ss}ler}, B., {Bamford}, S.~P., {Vika}, M., {et~al.} 2013, \mnras,
  430, 330

\bibitem[{{Head} {et~al.}(2014){Head}, {Lucey}, {Hudson}, \&
  {Smith}}]{Head2014}
{Head}, J.~T.~C.~G., {Lucey}, J.~R., {Hudson}, M.~J., \& {Smith}, R.~J. 2014,
  ArXiv e-prints

\bibitem[{{Hudson} {et~al.}(2010){Hudson}, {Stevenson}, {Smith}, {Wegner},
  {Lucey}, \& {Simard}}]{Hudson2010}
{Hudson}, M.~J., {Stevenson}, J.~B., {Smith}, R.~J., {et~al.} 2010, \mnras,
  409, 405

\bibitem[{{Jablonka} {et~al.}(1990){Jablonka}, {Alloin}, \&
  {Bica}}]{jablonka1990}
{Jablonka}, P., {Alloin}, D., \& {Bica}, E. 1990, \aap, 235, 22

\bibitem[{{Jaff{\'e}} {et~al.}(2011){Jaff{\'e}}, {Arag{\'o}n-Salamanca},
  {Kuntschner}, {Bamford}, {Hoyos}, {De Lucia}, {Halliday}, {Milvang-Jensen},
  {Poggianti}, {Rudnick}, {Saglia}, {Sanchez-Blazquez}, \&
  {Zaritsky}}]{Jaffe2011}
{Jaff{\'e}}, Y.~L., {Arag{\'o}n-Salamanca}, A., {Kuntschner}, H., {et~al.}
  2011, \mnras, 417, 1996

\bibitem[{{Jansen} {et~al.}(2000){Jansen}, {Franx}, {Fabricant}, \&
  {Caldwell}}]{Jansen2000}
{Jansen}, R.~A., {Franx}, M., {Fabricant}, D., \& {Caldwell}, N. 2000, \apjs,
  126, 271

\bibitem[{{Just} {et~al.}(2010){Just}, {Zaritsky}, {Sand}, {Desai}, \&
  {Rudnick}}]{Just2010}
{Just}, D.~W., {Zaritsky}, D., {Sand}, D.~J., {Desai}, V., \& {Rudnick}, G.
  2010, \apj, 711, 192

\bibitem[{{Kennicutt}(1998)}]{Kennicutt1998}
{Kennicutt}, Jr., R.~C. 1998, \araa, 36, 189

\bibitem[{{Koopmann} {et~al.}(2006){Koopmann}, {Haynes}, \&
  {Catinella}}]{Koopmann2006}
{Koopmann}, R.~A., {Haynes}, M.~P., \& {Catinella}, B. 2006, \aj, 131, 716

\bibitem[{{Koopmann} \& {Kenney}(2004)}]{Koopmann2004}
{Koopmann}, R.~A. \& {Kenney}, J.~D.~P. 2004, \apj, 613, 851

\bibitem[{{Magain} {et~al.}(1998){Magain}, {Courbin}, \& {Sohy}}]{MCS}
{Magain}, P., {Courbin}, F., \& {Sohy}, S. 1998, \apj, 494, 472

\bibitem[{{Mahajan} {et~al.}(2011){Mahajan}, {Mamon}, \&
  {Raychaudhury}}]{Mahajan2011}
{Mahajan}, S., {Mamon}, G.~A., \& {Raychaudhury}, S. 2011, \mnras, 416, 2882

\bibitem[{{Maltby} {et~al.}(2012){Maltby}, {Gray}, {Arag{\'o}n-Salamanca},
  {Wolf}, {Bell}, {Jogee}, {H{\"a}u{\ss}ler}, {Barazza}, {B{\"o}hm}, \&
  {Jahnke}}]{Maltby2012}
{Maltby}, D.~T., {Gray}, M.~E., {Arag{\'o}n-Salamanca}, A., {et~al.} 2012,
  \mnras, 419, 669

\bibitem[{{Maraston}(1998)}]{maraston1998}
{Maraston}, C. 1998, \mnras, 300, 872

\bibitem[{{Maraston} {et~al.}(2003){Maraston}, {Greggio}, {Renzini},
  {Ortolani}, {Saglia}, {Puzia}, \& {Kissler-Patig}}]{maraston2003}
{Maraston}, C., {Greggio}, L., {Renzini}, A., {et~al.} 2003, \aap, 400, 823

\bibitem[{{Masters} {et~al.}(2010){Masters}, {Mosleh}, {Romer}, {Nichol},
  {Bamford}, {Schawinski}, {Lintott}, {Andreescu}, {Campbell}, {Crowcroft},
  {Doyle}, {Edmondson}, {Murray}, {Raddick}, {Slosar}, {Szalay}, \&
  {Vandenberg}}]{Masters2010}
{Masters}, K.~L., {Mosleh}, M., {Romer}, A.~K., {et~al.} 2010, \mnras, 405, 783

\bibitem[{{Milvang-Jensen} {et~al.}(2008){Milvang-Jensen}, {Noll}, {Halliday},
  {Poggianti}, {Jablonka}, {Arag{\'o}n-Salamanca}, {Saglia}, {Nowak}, {von der
  Linden}, {De Lucia}, {Pell{\'o}}, {Moustakas}, {Poirier}, {Bamford}, {Clowe},
  {Dalcanton}, {Rudnick}, {Simard}, {White}, \&
  {Zaritsky}}]{Milvang-JensenB.2008}
{Milvang-Jensen}, B., {Noll}, S., {Halliday}, C., {et~al.} 2008, \aap, 482, 419

\bibitem[{{Moran} {et~al.}(2006){Moran}, {Ellis}, {Treu}, {Salim}, {Rich},
  {Smith}, \& {Kneib}}]{Moran2006}
{Moran}, S.~M., {Ellis}, R.~S., {Treu}, T., {et~al.} 2006, \apjl, 641, L97

\bibitem[{{Moustakas} {et~al.}(2011){Moustakas}, {Zaritsky}, {Brown}, {Cool},
  {Dey}, {Eisenstein}, {Gonzalez}, {Jannuzi}, {Jones}, {Kochanek}, {Murray}, \&
  {Wild}}]{Moustakas2011}
{Moustakas}, J., {Zaritsky}, D., {Brown}, M., {et~al.} 2011, ArXiv e-prints

\bibitem[{{Peng} {et~al.}(2011){Peng}, {Ho}, {Impey}, \& {Rix}}]{Peng2011}
{Peng}, C.~Y., {Ho}, L.~C., {Impey}, C.~D., \& {Rix}, H.-W. 2011, {GALFIT:
  Detailed Structural Decomposition of Galaxy Images}, astrophysics Source Code
  Library

\bibitem[{{Poggianti} {et~al.}(2008){Poggianti}, {Desai}, {Finn}, {Bamford},
  {De Lucia}, {Varela}, {Arag{\'o}n-Salamanca}, {Halliday}, {Noll}, {Saglia},
  {Zaritsky}, {Best}, {Clowe}, {Milvang-Jensen}, {Jablonka}, {Pell{\'o}},
  {Rudnick}, {Simard}, {von der Linden}, \& {White}}]{Poggianti2008}
{Poggianti}, B.~M., {Desai}, V., {Finn}, R., {et~al.} 2008, \apj, 684, 888

\bibitem[{{Poggianti} {et~al.}(1999){Poggianti}, {Smail}, {Dressler}, {Couch},
  {Barger}, {Butcher}, {Ellis}, \& {Oemler}}]{Poggianti1999}
{Poggianti}, B.~M., {Smail}, I., {Dressler}, A., {et~al.} 1999, \apj, 518, 576

\bibitem[{{Poggianti} {et~al.}(2006){Poggianti}, {von der Linden}, {De Lucia},
  {Desai}, {Simard}, {Halliday}, {Arag{\'o}n-Salamanca}, {Bower}, {Varela},
  {Best}, {Clowe}, {Dalcanton}, {Jablonka}, {Milvang-Jensen}, {Pello},
  {Rudnick}, {Saglia}, {White}, \& {Zaritsky}}]{Poggianti2006}
{Poggianti}, B.~M., {von der Linden}, A., {De Lucia}, G., {et~al.} 2006, \apj,
  642, 188

\bibitem[{{Rodr{\'{\i}}guez Del Pino} {et~al.}(2014){Rodr{\'{\i}}guez Del
  Pino}, {Bamford}, {Arag{\'o}n-Salamanca}, {Milvang-Jensen}, {Merrifield}, \&
  {Balcells}}]{Delpino2014}
{Rodr{\'{\i}}guez Del Pino}, B., {Bamford}, S.~P., {Arag{\'o}n-Salamanca}, A.,
  {et~al.} 2014, \mnras, 438, 1038

\bibitem[{{Rudnick} {et~al.}(2006){Rudnick}, {Labb{\'e}}, {F{\"o}rster
  Schreiber}, {Wuyts}, {Franx}, {Finlator}, {Kriek}, {Moorwood}, {Rix},
  {R{\"o}ttgering}, {Trujillo}, {van der Wel}, {van der Werf}, \& {van
  Dokkum}}]{Rudnick2006}
{Rudnick}, G., {Labb{\'e}}, I., {F{\"o}rster Schreiber}, N.~M., {et~al.} 2006,
  \apj, 650, 624

\bibitem[{{Rudnick} {et~al.}(2009){Rudnick}, {von der Linden}, {Pell{\'o}},
  {Arag{\'o}n-Salamanca}, {Marchesini}, {Clowe}, {De Lucia}, {Halliday},
  {Jablonka}, {Milvang-Jensen}, {Poggianti}, {Saglia}, {Simard}, {White}, \&
  {Zaritsky}}]{Rudnick2009}
{Rudnick}, G., {von der Linden}, A., {Pell{\'o}}, R., {et~al.} 2009, \apj, 700,
  1559

\bibitem[{{Simard} {et~al.}(2009){Simard}, {Clowe}, {Desai}, {Dalcanton}, {von
  der Linden}, {Poggianti}, {White}, {Arag{\'o}n-Salamanca}, {De Lucia},
  {Halliday}, {Jablonka}, {Milvang-Jensen}, {Saglia}, {Pell{\'o}}, {Rudnick},
  \& {Zaritsky}}]{Simard2009}
{Simard}, L., {Clowe}, D., {Desai}, V., {et~al.} 2009, \aap, 508, 1141

\bibitem[{{Simard} {et~al.}(2002){Simard}, {Willmer}, {Vogt}, {Sarajedini},
  {Phillips}, {Weiner}, {Koo}, {Im}, {Illingworth}, \& {Faber}}]{Simard2002}
{Simard}, L., {Willmer}, C.~N.~A., {Vogt}, N.~P., {et~al.} 2002, \apjs, 142, 1

\bibitem[{{Taranu} {et~al.}(2014){Taranu}, {Hudson}, {Balogh}, {Smith},
  {Power}, {Oman}, \& {Krane}}]{Taranu2014}
{Taranu}, D.~S., {Hudson}, M.~J., {Balogh}, M.~L., {et~al.} 2014, \mnras, 440,
  1934

\bibitem[{{Vika} {et~al.}(2013){Vika}, {Bamford}, {H{\"a}u{\ss}ler}, {Rojas},
  {Borch}, \& {Nichol}}]{Vika2013}
{Vika}, M., {Bamford}, S.~P., {H{\"a}u{\ss}ler}, B., {et~al.} 2013, \mnras,
  435, 623

\bibitem[{{Villalobos} {et~al.}(2012){Villalobos}, {De Lucia}, {Borgani}, \&
  {Murante}}]{Villalobos2012}
{Villalobos}, {\'A}., {De Lucia}, G., {Borgani}, S., \& {Murante}, G. 2012,
  \mnras, 424, 2401

\bibitem[{{White} {et~al.}(2005){White}, {Clowe}, {Simard}, {Rudnick}, {De
  Lucia}, {Arag{\'o}n-Salamanca}, {Bender}, {Best}, {Bremer}, {Charlot},
  {Dalcanton}, {Dantel}, {Desai}, {Fort}, {Halliday}, {Jablonka}, {Kauffmann},
  {Mellier}, {Milvang-Jensen}, {Pell{\'o}}, {Poggianti}, {Poirier},
  {Rottgering}, {Saglia}, {Schneider}, \& {Zaritsky}}]{White2005}
{White}, S.~D.~M., {Clowe}, D.~I., {Simard}, L., {et~al.} 2005, \aap, 444, 365

\bibitem[{{Williams} {et~al.}(2009){Williams}, {Quadri}, {Franx}, {van Dokkum},
  \& {Labb{\'e}}}]{williams2009}
{Williams}, R.~J., {Quadri}, R.~F., {Franx}, M., {van Dokkum}, P., \&
  {Labb{\'e}}, I. 2009, \apj, 691, 1879

\bibitem[{{Wolf} {et~al.}(2009){Wolf}, {Arag{\'o}n-Salamanca}, {Balogh},
  {Barden}, {Bell}, {Gray}, {Peng}, {Bacon}, {Barazza}, {B{\"o}hm}, {Caldwell},
  {Gallazzi}, {H{\"a}u{\ss}ler}, {Heymans}, {Jahnke}, {Jogee}, {van Kampen},
  {Lane}, {McIntosh}, {Meisenheimer}, {Papovich}, {S{\'a}nchez}, {Taylor},
  {Wisotzki}, \& {Zheng}}]{Wolf2009}
{Wolf}, C., {Arag{\'o}n-Salamanca}, A., {Balogh}, M., {et~al.} 2009, \mnras,
  393, 1302

\bibitem[{{Woo} {et~al.}(2015){Woo}, {Dekel}, {Faber}, \& {Koo}}]{Woo2015}
{Woo}, J., {Dekel}, A., {Faber}, S.~M., \& {Koo}, D.~C. 2015, \mnras, 448, 237

\bibitem[{{Worthey} \& {Ottaviani}(1997)}]{Worthey1997}
{Worthey}, G. \& {Ottaviani}, D.~L. 1997, \apjs, 111, 377

\end{thebibliography}
\end{document}